\documentclass[fleqn,usenatbib]{mnras}

\usepackage{newtxtext,newtxmath}

\usepackage[T1]{fontenc}

\usepackage{graphicx}
\usepackage{multirow}
\usepackage{xcolor}
\usepackage{amsmath}
\usepackage{orcidlink}

\title[Mapping local luminosity-density anisotropies]
{Everything Everywhere All At Once: A Hierarchical Framework for Mapping Anisotropies in the Local Universe}

\author[A. Bansal et al.]{
Aryan Bansal$^{1}$\orcidlink{0000-0002-6631-1886}\thanks{E-mail: aryanbansal@swin.edu.au},
Edward N. Taylor$^{1}$\orcidlink{0000-0002-5522-9107},
Michelle Cluver$^{1}$\orcidlink{0000-0002-9871-6490},
and Matthew Colless$^{2}$\orcidlink{0000-0001-9552-8075}
\\
$^{1}$Centre for Astrophysics and Supercomputing, Swinburne University of Technology, Hawthorn, VIC 3122, Australia\\
$^{2}$Research School of Astronomy and Astrophysics, Australian National University, Canberra, ACT 2611, Australia
}

\date{Accepted XXX. Received YYY; in original form ZZZ}

\pubyear{\the\year{}}

\begin{document}
\label{firstpage}
\pagerange{\pageref{firstpage}--\pageref{lastpage}}
\maketitle

\begin{abstract}
We present a Bayesian hierarchical forward model for mapping the local luminosity-density field from heterogeneous redshift surveys. By dividing the survey volume into angular cells and redshift shells, we model the observed galaxy distribution directly in apparent-magnitude and redshift space. This allows us to infer cell-level luminosity functions and densities while preserving each survey's unique selection function. Simultaneously, the ensemble of cells constrains the cosmic-mean luminosity function, its redshift evolution, and the hyperparameters describing cell-to-cell variation. We apply the framework to 2MRS, 6dFGS, and GAMA over
$0.005<z<0.065$, combining wide sky coverage with deeper and
fainter galaxy samples. The model successfully recovers global $J$-band luminosity-function parameters and a mean luminosity density consistent with previous low-redshift measurements. The inferred luminosity-density scatter decreases with volume, consistent with the cosmic-variance amplitude expected in a $\Lambda$CDM universe. Within the volume we probe, we find no evidence for a large coherent underdensity. Instead, the local Universe is broadly consistent with smooth evolution and cosmic variance. The nearest shell at $z \simeq 0.01$ is the clearest exception, lying $\simeq 0.12$~dex ($\simeq 25\%$) below the smooth global model, a $-2.6\sigma$ underdensity with a one-sided significance of $\simeq 99.5\%$. This framework offers a scalable architecture for mapping cosmic density fields with next-generation wide and deep surveys.
\end{abstract}




\section{Introduction}
\label{sec:intro}

The cosmological principle underlies the standard description of the Universe on large scales. 
It states that, when averaged over sufficiently large volumes, the Universe is statistically homogeneous and isotropic \citep{Peebles1980,Peebles1993}. 
This assumption is built into the Friedmann--Lema\^{i}tre--Robertson--Walker metric and forms the basis of the $\Lambda$CDM model \citep{Dodelson2003,Weinberg2008}. 
Measurements of the Cosmic Microwave Background show temperature fluctuations at the level of $\Delta T/T \sim 10^{-5}$, providing strong evidence for statistical isotropy at early times \citep{PlanckCollaboration2020}. 
These small initial fluctuations are amplified over cosmic time by gravitational collapse, giving rise to the late-time structure observed in the galaxy distribution.
At low redshift, therefore, the situation is more complicated. 
The galaxy distribution is highly structured, and any measurement of homogeneity must be made using surveys with finite depth, incomplete sky coverage, and non-trivial selection functions.

Galaxies trace the large-scale structure of the Universe, revealing a hierarchy of voids, sheets, filaments, and clusters. 
Statistical analyses indicate that the galaxy distribution approaches homogeneity on scales of order $\sim70-100\,h^{-1}\,\mathrm{Mpc}$, although the inferred scale depends on tracer population and methodology. 
Counts-in-spheres analyses have been applied to SDSS luminous red galaxies and WiggleZ galaxies \citep{Hogg2005,Scrimgeour2012}, while related studies have used information entropy, multifractal statistics, and corrected counts-in-spheres measurements \citep{Sarkar2009,Nadathur2013}. 
Alternative analyses based on conditional densities and correlation functions have argued for larger homogeneity scales, highlighting the sensitivity of the result to methodology \citep{SylosLabini2009}.  
Even if homogeneity holds statistically on large scales, low-redshift measurements probe limited cosmological volumes and are therefore strongly affected by cosmic variance \citep{Peebles1980,Driver2010}. 
Several studies have reported large density variations in the nearby Universe, including early counts-in-cells measurements \citep{1990Natur.348..705E,Efstathiou1995,Loveday1992_PaperI} and suggestions of a large-scale underdensity extending to $z\sim0.1$--$0.2$ \citep{Keenan2013,Busswell2004,Wong2022,Shanks2019,WhitbournShanks2014,2016MNRAS.459..496W}.
Such structures are not necessarily in conflict with $\Lambda$CDM, but they show that local measurements require a method that can separate real spatial variation from survey geometry, selection effects, and smooth redshift evolution. 
Coherent peculiar-velocity flows provide additional evidence of significant inhomogeneities even on scales as large as $200-300$ Mpc; and that the nearby Universe cannot simply be treated as uniform on these scales \citep{Watkins2009}.

Complementary probes have also been used to study large-scale structure. 
Baryon acoustic oscillation measurements trace clustering on large scales \citep{Eisenstein2005,Cole2005,Anderson2014}, Quasar samples have been used to extend isotropy tests to higher redshifts \citep{Secrest2021}, and radio surveys have been used to measure dipole amplitudes \citep{RubartSchwarz2013}. 
These approaches probe different aspects of large-scale structure, but generally rely on number counts, standard candles, or clustering statistics.

These issues are relevant to current tensions and isotropy tests in observational cosmology. 
The discrepancy between early- and late-Universe measurements of the Hubble constant remains unresolved \citep{Verde2019,Riess2022,2024ApJ...962L..17R,2025ApJ...978L..33G,2025ApJ...985..203F}. 
Local density fluctuations can contribute to shifts in the locally inferred expansion rate \citep{Keenan2013,Wong2022,Freedman2021}. 
Beyond the Hubble constant, several probes have reported hints of large-scale anisotropy, including measurements of the radio dipole amplitude \citep{RubartSchwarz2013} and isotropy tests using quasars \citep{Secrest2021}. 
These results motivate analyses that measure spatial variation directly, while also distinguishing genuine structure from fluctuations expected within $\Lambda$CDM.

The difficulty is not only that the local Universe is structured, but also that the available data are fragmented. 
No single low-redshift survey provides all of the sky coverage, depth, completeness, and uniformity needed to map the luminosity-density field on its own. 
Wide-area surveys are essential for measuring large-scale structure, but are usually shallow \citep{2012ApJS..199...26H, 2004MNRAS.355..747J, 2002AJ....124.1810S}. 
Deeper surveys constrain the faint galaxy population more effectively, but cover only small regions of sky \citep{2011MNRAS.413..971D}. 
In addition, each survey has its own magnitude limits, footprint, completeness, masks, and calibration choices (as discussed in value-added compilations like \cite{2005AJ....129.2562B}). 
A model designed for the local Universe must therefore do more than fit one average luminosity function. 
It must combine different surveys, preserve their individual selection functions, allow spatial variation, and still remain computationally feasible.

In this work we use the galaxy luminosity function (LF) as the link between the observed galaxy distribution and the luminosity-density field. 
The LF describes the number density of galaxies as a function of luminosity and is well represented by the Schechter form \citep{Schechter1976}. 
Measurements of the LF require careful treatment of selection effects and incompleteness \citep{Efstathiou1988}, and at low redshift are sensitive to cosmic variance \citep{Driver2010}. 
Simple number counts provide a direct measure of galaxy density, but they are strongly affected by survey depth, selection limits, and the relative contribution of faint galaxies. 
They are therefore difficult to compare consistently across surveys and redshift ranges without a model for the underlying galaxy population. 
If the LF can be inferred, a more physically useful quantity is the luminosity density,
\begin{equation}
\begin{aligned}
j(z) &= \int_{0}^{\infty}
L\,\phi(L,z)\,\mathrm{d}L .
\end{aligned}
\label{eq:luminosity_density}
\end{equation}
Here, $\phi(L,z)$ is the redshift-dependent luminosity function expressed in terms of luminosity.
The luminosity corresponding to absolute magnitude $M$ is
\begin{equation}
\begin{aligned}
L(M) &= L_{\odot,J}\,
10^{-0.4\left(M-M_{\odot,J}\right)} .
\end{aligned}
\label{eq:luminosity}
\end{equation}
The luminosity density measures the total emitted light per unit volume. 
It provides a proxy for the stellar-mass density and, modulo galaxy bias, traces the mean matter density within a volume. 
It is therefore a useful quantity for studying coherent over- and under-densities in the local Universe.

We develop a Bayesian hierarchical framework for measuring this luminosity-density field from heterogeneous redshift surveys. 
The survey volume is divided into angular cells and redshift shells, and each region is assigned its own LF parameters. 
These cell-level parameters are not fitted independently; they are linked through shared hyperparameters that describe the cosmic mean and the intrinsic scatter about that mean. 
This allows information to be shared across regions and surveys, while still allowing individual cells to deviate from the mean where the data support it.

The key idea is to treat the full problem in one forward model. 
Rather than correcting the data into a simplified form and then fitting separate pieces afterwards, we model the observed galaxy distribution directly in apparent-magnitude and redshift space. 
Selection limits, effective areas, number counts, LF parameters, luminosity density, and luminosity-density scatter are all connected through the same likelihood. 
This is the sense in which the model fits ``everything'', works ``everywhere'' and ``all at once'': the same statistical structure is applied across sky position, redshift, and survey overlap, and the parameters are inferred jointly rather than in a sequence of separate steps. 
The title is therefore not just a joke, but a description of the modelling strategy. 
The framework is built to keep the relevant pieces of the problem together: the galaxies, the surveys, the luminosity functions, the density field, and the scatter around the cosmic mean.

The cosmological application in this paper provides a demanding test of the framework. 
If the model can combine current low-redshift surveys, recover a sensible global luminosity function, map local luminosity-density variations, and measure scatter consistent with cosmic variance, then it provides a useful baseline for the next generation of wide and deep surveys. 
The present data are not perfect for testing the largest proposed local underdensities, but they are an excellent testbed for the method. 
Future surveys will make this type of hierarchical luminosity-density mapping substantially more powerful.

The paper is organised as follows. 
Section~\ref{sec:data} describes the survey data and selection criteria. 
Section~\ref{sec:model} presents the hierarchical LF model and computational implementation. 
Section~\ref{sec:results} presents the inferred LF evolution and luminosity density maps. 
Section~\ref{sec:discussion} discusses implications for local inhomogeneity within the standard cosmological framework and summarises the conclusions.\\

\textbf{Conventions and modelling choices:}
We adopt a flat $\Lambda$CDM cosmology with $H_0=70~\mathrm{km\,s^{-1}\,Mpc^{-1}}$ and $\Omega_m=0.3$. 
Magnitudes and luminosities are reported in the 2MASS $J$-band Vega system unless otherwise stated, and should be understood as observer-frame quantities under the bandpass treatment described in Section~\ref{sec:model}. 
We do not apply explicit galaxy-by-galaxy $K$-corrections in the present analysis; this modelling choice is discussed in Section~\ref{sec:model}. 
We convert from AB to Vega magnitudes using the fixed zero-point offset \citep{Cohen2003}:
\begin{equation}
J^{\rm Vega} = J^{\rm AB} - 0.916 .
\label{eq:ABtoVega}
\end{equation}

\section{Data}
\label{sec:data}

In order to measure cosmic variance and test the consistency of the cosmological principle in the local Universe, we require a description of the galaxy distribution over the widest possible volume while retaining sufficient depth and statistical power to constrain the galaxy luminosity function (LF). For this purpose, we combine spectroscopic redshift surveys with near-infrared $J$-band photometry. The 2MASS $J$-band data are particularly suitable for this analysis because it is well callibrated, uniform across the sky, is less sensitive to dust extinction than optical bands, and provides a more direct tracer of the underlying stellar mass density.

\begin{figure*}
    \centering
    
    \includegraphics[width=0.8\linewidth]{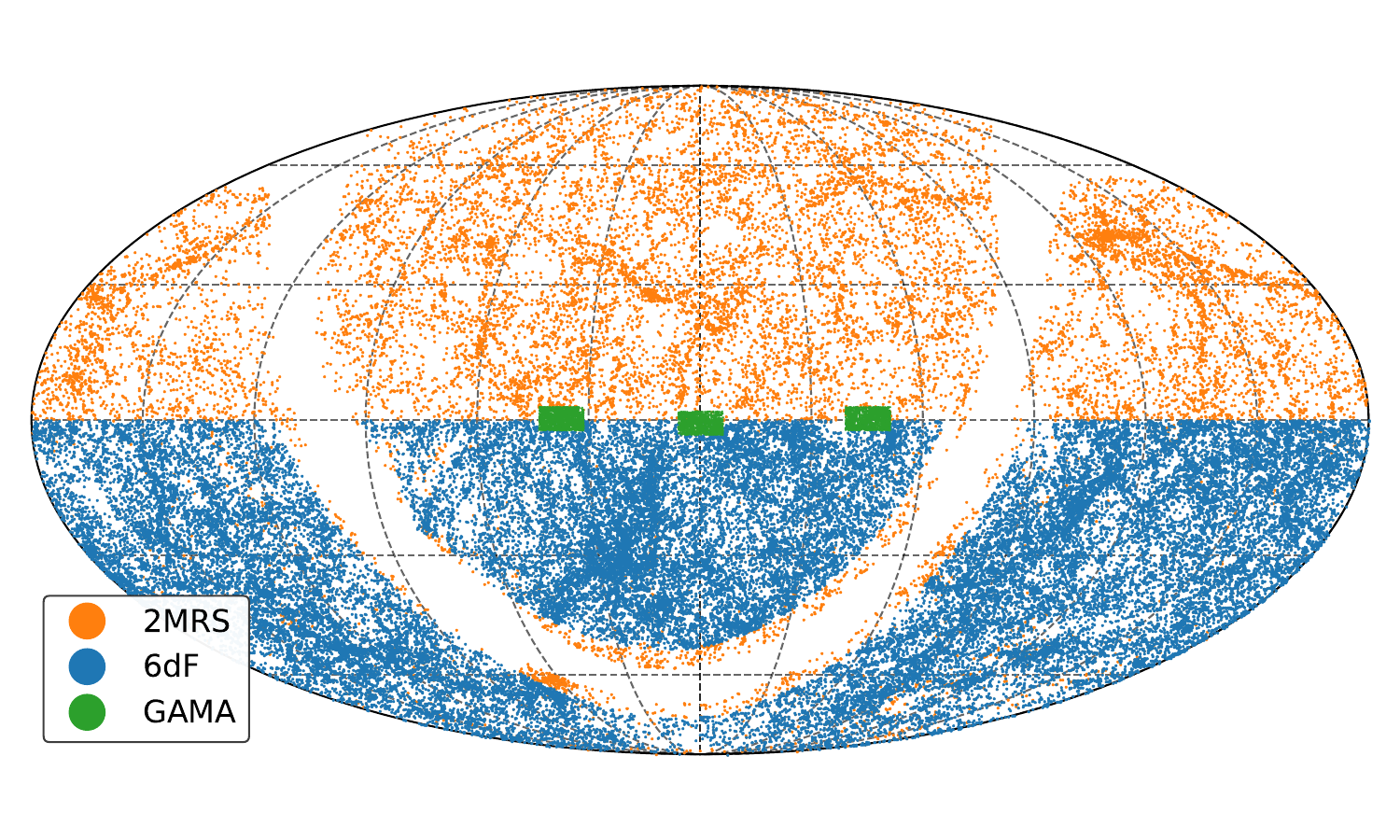}
    \caption{Sky coverage of the 2MRS, 6dF, and GAMA surveys used in this work.}
    \label{healpix_surveys}
\end{figure*}

\begin{figure*}
    \includegraphics[width=1.0\linewidth]{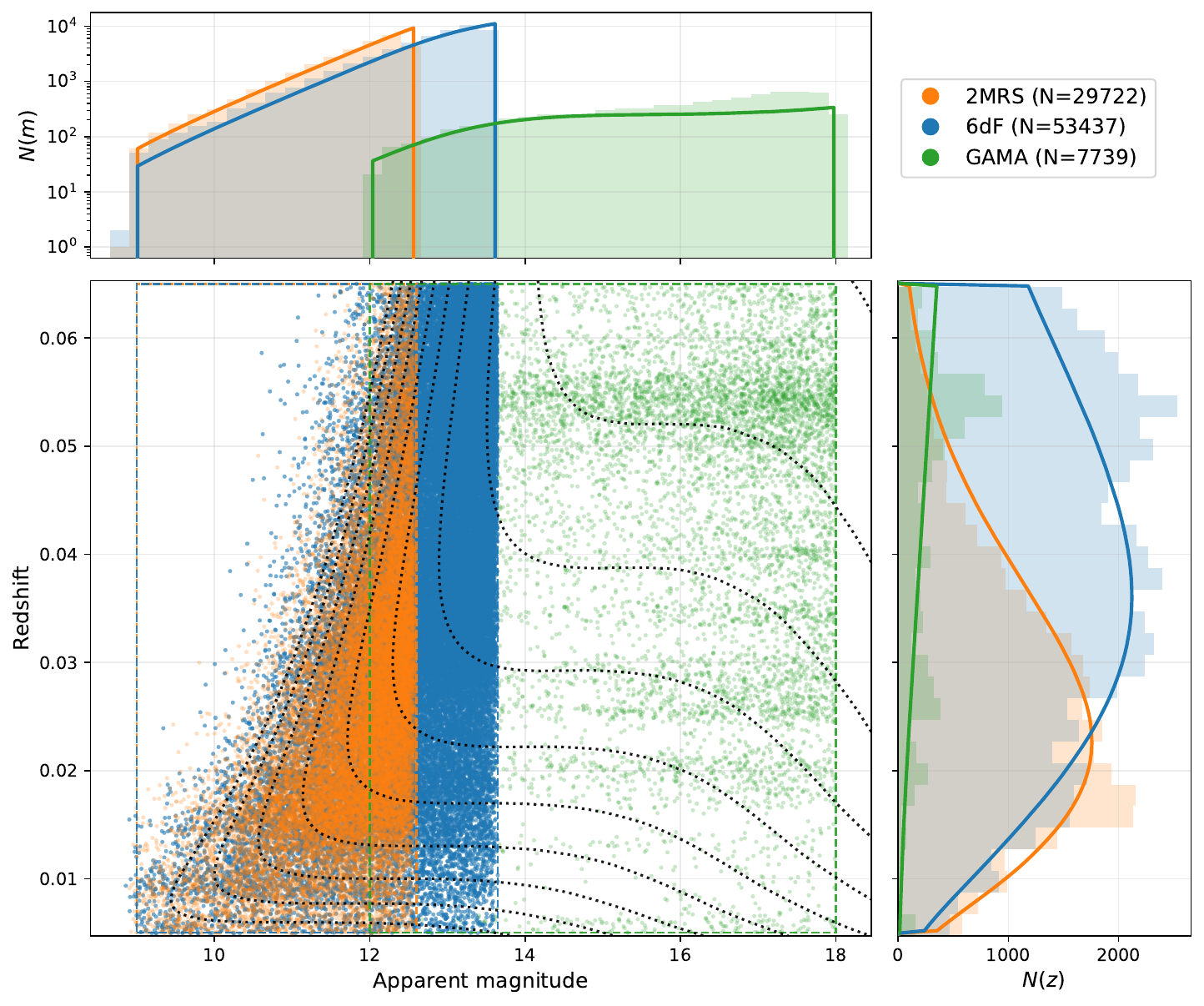}
    \caption{
    Survey magnitude and redshift limits applied to the data. 
    2MRS galaxies are shown in orange, 6dF in blue, and GAMA in green. 
    Centre: distribution of galaxies in apparent magnitude--redshift $(m,z)$ space. 
    The dotted curves show representative equal-density contours predicted by the luminosity-function model. 
    Top: projected apparent-magnitude distributions $N(m)$ for each
    survey. The histograms show the observed galaxy counts, while the
    solid curves show the prediction from the same global model used
    for all three surveys, rather than independent fits to each survey.
    The faint-end excess in GAMA is associated mainly with the overdense
    structure near $z\simeq0.06$ visible in the right panel.
    Right: projected redshift distributions $N(z)$ for each survey, with histograms showing the observed galaxy counts and solid curves showing the fitted model predictions.
    }
    \label{survey_limits}
\end{figure*}

\subsection{Surveys}
We use three complementary datasets: the Two Micron All-Sky Redshift Survey (2MRS), which provides highly complete all-sky coverage at low redshift; the 6dF Galaxy Survey (6dFGS), which reaches greater depth across most of the southern sky; and the Galaxy And Mass Assembly survey (GAMA), which reaches much fainter galaxies over a smaller area. Their sky coverage is shown in Figure~\ref{healpix_surveys}.
As discussed in Section~\ref{sec:model}, our hierarchical framework combines these surveys self-consistently, allowing each survey to constrain different regions of the luminosity function while compensating for the limitations of the others.

The 2MASS Redshift Survey (2MRS; \citealt{2012ApJS..199...26H}) is selected from the 2MASS Extended Source Catalogue (XSC) \citep{2006AJ....131.1163S} and achieves a completeness of approximately $97\%$ to a limiting magnitude of $J = 12.70$. The survey has a median redshift of $z_{\rm med}\simeq0.028$ and provides near all-sky coverage over approximately $37{,}540~\mathrm{deg}^2$. Owing to its high completeness and wide angular coverage, 2MRS is useful for constraining the number counts of massive nearby galaxies, but not deep enough for constaining the faint-end of the luminosity function on its own.

The 6dF Galaxy Survey (6dFGS (hereafter 6dF); \citealt{2004MNRAS.355..747J, 2009MNRAS.399..683J}) is also selected from the 2MASS XSC and covers most of the southern sky to a limiting magnitude of $J = 13.65$. The survey completeness is approximately $88\%$, with a median redshift of $z_{\rm med}\simeq0.05$, over a footprint of approximately $17{,}000~\mathrm{deg}^2$. Relative to 2MRS, 6dF probes significantly deeper magnitudes for better statistics, but covers only the southern sky..

The Galaxy And Mass Assembly survey (GAMA; \citealt{2015MNRAS.452.2087L, 2011MNRAS.413..971D} is a deep panchromatic survey using VISTA imaging \citep{2020MNRAS.496.3235B}. Unlike 2MRS and 6dF, GAMA covers a much smaller region of sky, approximately $286~\mathrm{deg}^2$, but achieves very high spectroscopic completeness of approximately $98\%$ to $J \simeq 18$; which is around $50$x deeper than 6dF in luminosity. The survey reaches substantially fainter galaxies and higher redshifts, with $z_{\rm med}\simeq0.15$, making it particularly important for constraining the faint-end of the luminosity function, especially the faint-end slopes. GAMA photometry is provided in the AB magnitude system. To maintain consistency with the 2MASS Vega system used by 2MRS and 6dF, we have converted the AB magnitudes given by GAMA to the Vega system using equation \ref{eq:ABtoVega}.

\subsection{Sample Selection}
We impose fixed apparent-magnitude limits for each survey and use a common redshift range for the present analysis. For 6dF we select galaxies with $9.00 \le J \le 13.65$; for 2MRS, $9.00 \le J \le 12.70$; and for GAMA, we use the Data Release 4 (DR4) \citep{2022MNRAS.513..439D} data within the three equatorial fields (G09, G12 and G15) spanning $12.00 \le J \le 18.00$. In addition to the faint limits set by survey completeness, we impose bright magnitude limits in order to exclude very nearby extended galaxies whose photometry is more strongly affected by saturation and surface-brightness systematics. We further restrict the analysis to the redshift range
\begin{equation}
0.005 < z < 0.065 .
\end{equation}
The lower limit is imposed to remove residual stellar contamination and very nearby sources, while the upper limit corresponds to the regime where the wide-area surveys begin to become sparse. These limits define the survey selection functions used in the likelihood normalisation described in Section~\ref{sec:lf_forward_model}. The adopted limits are illustrated in Figure~\ref{survey_limits}. The framework does not require all surveys to share the same
redshift range. We adopt a common range here because our aim is to
map the luminosity-density field within the local Universe. GAMA is
therefore used for its faint-end leverage within this volume, rather
than for its full redshift depth.

\subsection{Data Corrections}

The survey footprints and effective areas are incorporated through random catalogues. 
For 6dF we use the random catalogue of \citet{2011MNRAS.416.3017B}, while for 2MRS we construct a random catalogue directly from the survey mask. 
These random catalogues are used to estimate the effective area of each spatial cell, which enters the likelihood normalisation described in Section~\ref{sec:model}. 
No additional spectroscopic-completeness correction is applied to 2MRS or GAMA, since both surveys are already highly complete over the adopted selection limits.
The 6dF random catalogue constrains the relative angular completeness of 6dF across the sky, but does not fully determine the global completeness of 6dF relative to 2MRS. 
We therefore include a multiplicative scaling factor, $k_{\rm 6dF}$, to account for the residual relative completeness of 6dF with respect to 2MRS, as discussed further in Section~\ref{sec:model_choices}.

Galactic extinction corrections are applied using \citet{Schlegel1998} with the recalibration of \citet{SchlaflyFinkbeiner2011}. 
Because the model is defined in observed apparent-magnitude space, foreground dust is treated as a correction to the effective magnitude limits in each spatial cell, rather than as a galaxy-by-galaxy correction to the luminosity function. 
The details of this per-cell dust treatment and its impact on the expected number counts are given in Appendix~\ref{app:dust}.

All redshifts are converted from the heliocentric frame to the CMB frame using the dipole correction of \citet{Kogut1993}. 
Bulk-flow corrections are then applied using the maps of \citet{2020MNRAS.497.1275S,2022ApJ...938..112P,2022PASA...39...46C} and the \texttt{PVhub} Python package \citep{said_pvhub_2024}. 
We do not apply group-based velocity-dispersion corrections for the fingers-of-god effect, because existing group catalogues are only available for a subset of the 6dF sample and cannot be applied consistently across all surveys. 
We nevertheless test the possible impact of fingers-of-god corrections in several nearby clusters with strong velocity dispersions, and find that the inferred luminosity density is not significantly affected. 
These tests are presented in Appendix~\ref{appendix:fog}.

\section{Model}
\label{sec:model}

In this section, we describe the model used to infer spatial variations in the galaxy luminosity function and luminosity density. 
The main goal is to measure how the luminosity density varies across the local Universe. 
We do this by dividing the survey volume into angular cells and redshift shells, using a Bayesian hierarchical model to fit the galaxy luminosity function in each spatial region, and then integrating the fitted luminosity function to obtain the luminosity density. 

This section describes and motivates the most important features of our modelling approach. 
A more formal definition of the model can be found in Appendix~\ref{app:model}, including several important implementation details.

\subsection{Luminosity Function and Forward Model}
\label{sec:lf_forward_model}

The luminosity function, $\phi(M,z)$, is a counting function for galaxies. 
It describes the differential number density of galaxies as a function of absolute magnitude and redshift. 
For a complete, volume-limited sample, estimating the luminosity function would be straightforward: one could simply count galaxies in bins of absolute magnitude and divide by the survey volume,
\begin{equation}
\phi(M,z)\,dM\,dz
\simeq
\frac{N(M,z)\,dM\,dz}{V}.
\end{equation}
In practice, however, the luminosity function is defined in absolute magnitude, while the directly observed quantities are apparent magnitude and redshift. The absolute magnitude must therefore be inferred from these observables, accounting for the distance modulus, bandpass shifting, luminosity evolution, and foreground extinction,
\begin{equation}
M =
m - DM(z) - K(z) - E(z) - A,
\label{eq:M-m}
\end{equation}
where $DM(z)$ is the distance modulus, $K(z)$ is the $K$-correction, $E(z)$ describes luminosity evolution, and $A$ is the Galactic extinction in the observed band. 
In the present forward model, Galactic extinction is incorporated at the level of the survey selection by shifting the effective apparent-magnitude limits in each spatial cell, rather than by correcting each galaxy individually. 
This keeps the selection function inside the likelihood normalisation and is described in Appendix~\ref{app:dust}.

Simple selection effects are traditionally handled by weighting each
galaxy by the volume over which it could have been observed. This
gives the classical $1/V_{\max}$ estimator
\citep{1968ApJ...151..393S},
\begin{equation}
\phi(M_j)\,dM
\approx
\sum_i \frac{1}{V_{{\rm max},i}},
\end{equation}
where the sum is over galaxies in a magnitude bin centred on $M_j$.
The accessible volume for each galaxy is
\begin{equation}
V_{{\rm max},i}
=
d\Omega
\int_0^{z_{{\rm max},i}}
\frac{dV}{dz}\,dz,
\end{equation}
where $d\Omega$ is the survey solid angle and $z_{{\rm max},i}$ is
the maximum redshift at which that galaxy would remain within the
survey selection limits.

The $1/V_{\max}$ estimator assumes a uniform spatial distribution of
galaxies within the accessible volume, and is therefore known to
become biased in the presence of large-scale structure
\citep{2018MNRAS.474.5500O}. This is a problem here because our goal
is precisely to measure spatial variation rather than average over it.

Likelihood-based estimators were developed to reduce this sensitivity to density inhomogeneities. The Step-Wise Maximum Likelihood method \citep[SWML;][]{Efstathiou1988} provides a non-parametric estimate of the luminosity function, while the \citet*{1979ApJ...232..352S} (hereafter STY) method fits a Schechter function directly to the galaxy distribution. \citet{cole2011} extended the SWML approach to fit the luminosity function and density fluctuations together, with the method developed further by \citet{Loveday2015_LF}.

Our approach is closer in spirit to the STY method, but extends the problem by allowing the luminosity
function itself to vary between spatial cells while jointly modelling
the galaxy distribution, number counts, and the different selection
functions of multiple surveys. Unlike standard SWML, which estimates
a stepwise luminosity function for the selected sample, we fit a
continuous parametric luminosity function directly to the observed
$(m,z)$ distribution and allow its mean parameters to evolve smoothly
with redshift. The cell-level parameters are then linked through a
common hierarchical model rather than fitted independently, allowing
redshift evolution and spatial variation to be treated within the same
likelihood.

We therefore constrain the luminosity function through a Bayesian
forward model. The starting point is that absolute magnitude is a
derived quantity. The direct observables are apparent magnitude and
redshift. Accordingly, we write the model in terms of the expected
galaxy distribution in observed space, $N(m,z)$, rather than first
constructing an estimator for $\phi(M,z)$ or $N(M,z)$.

If the luminosity function parameters are known, the expected number of galaxies in an observed element of apparent magnitude and redshift is
\begin{equation}
N(m,z)
=
d\Omega\,
\frac{dV}{dz}\,
\phi\big(m-\Delta_z,z\big),
\end{equation}
where we define
\begin{equation}
\Delta_z \equiv m-M .
\end{equation}
In the general case, $\Delta_z$ contains the redshift-dependent terms described in Equation~\ref{eq:M-m}. 
In this work, we use the baseline bandpass-stretching correction \citep{1999astro.ph..5116H,1968ApJ...154...21O,2003ApJ...592..819B}; the full magnitude relation is given in Appendix~\ref{app:model}.
Smooth mean redshift-dependent effects can be partly degenerate with the luminosity-function evolution parameters \citep{2003ApJ...592..819B}. Galaxy-to-galaxy SED-dependent effects are not included in this way, as discussed in Section~\ref{subsec:LF_param}.
Examples of the resulting forward-modelled distributions are shown in Figures~\ref{fig:gama_lf} and \ref{fig:cells_lf}.

The likelihood of observing a galaxy at a particular position in apparent-magnitude and redshift space is directly related to the bivariate density $N(m,z)$, up to a normalising constant. 
To convert this predicted density into a probability distribution, we normalise it over the same survey window used to compute the expected number of galaxies. 
The total expected number of galaxies is
\begin{equation}
N_{\rm pred}
=
\int dm \int dz\,N(m,z).
\label{eq:npred}
\end{equation}
Using the expression above, the single-galaxy likelihood contribution can therefore be written as
\begin{equation}
\mathcal{L}(m,z)
=
\frac{1}{N_{\rm pred}}\,
d\Omega\,
\frac{dV}{dz}\,
\phi\big(m-\Delta_z,z\big).
\end{equation}
This is the likelihood used to infer the luminosity-function parameters in the usual Bayesian way.

It is important to recognise that, relative to the standard absolute-magnitude approach, this reformulation is only a change of variables from $M$ to $m=M+\Delta_z$. 
The luminosity function is still defined in absolute magnitude, but is evaluated at the absolute magnitude implied by each observed pair $(m,z)$.

This formulation is useful because, for all galaxies within a given survey cell, the apparent-magnitude and redshift limits are fixed by the survey selection. 
The normalising integral in equation \ref{eq:npred}, therefore has the same integration limits for every galaxy in that cell, and only needs to be evaluated once per cell per model evaluation.
The model also remains continuous in redshift, since the likelihood is written in terms of the smooth bivariate density $N(m,z)$.

These points matter as the luminosity-function parameters
themselves are not the final target of this paper. They are the means
by which we infer the luminosity density in each cell. For a given
cell, the luminosity density is evaluated using
Equation~\ref{eq:luminosity_density}, with the cell-level luminosity
function inserted for $\phi(M,z)$. In this sense, the luminosity
function provides the local model of the galaxy population, while
$j$ provides the physical summary of the total amount of
$J$-band light per unit volume in that cell. 

With this forward model
established, the remainder of this section describes how we evaluate
the model in spatial cells, parameterise the luminosity function, and
infer the cell-level luminosity densities across the sky.

\subsection{Spatial Cells and Survey Selection}
\label{sec:model_cells}

To measure spatial variations, the survey volume is divided into angular and radial cells. 
Each cell is defined by an angular sky pixel and a redshift shell. 
For 2MRS and 6dF, we divide the sky using the \textsc{HEALPix}
pixelisation scheme with \texttt{nside}$=3$, giving 108 angular
pixels of approximately $382\,\deg^2$ each. We further divide the
redshift range $0.005<z<0.065$ into six shells of width
$\Delta z=0.01$. Each spatial cell is therefore defined by the
intersection of one angular pixel and one redshift shell, giving
$108\times6=648$ possible cells across the full sky. GAMA has a much smaller sky footprint and is used primarily to constrain the faint end of the luminosity function, so its three fields are treated together as a single region rather than as independent spatial cells.

The survey footprints are incomplete and irregular, so the effective
area of each cell is estimated using random catalogues. For each cell
the effective area is given by
\begin{align}
f_c &=
\frac{N_{{\rm rand},c}}
{N_{\rm rand,tot}},
\\
\Omega_{{\rm eff},c}
&=
f_c\,\Omega_{\rm survey}.
\end{align}
Here, $N_{{\rm rand},c}$ is the number of random points in the
angular pixel containing cell $c$, and $N_{\rm rand,tot}$ is the total
number of random points across the survey footprint. The ratio $f_c$
therefore gives the fraction of the survey area contained in that
pixel. $\Omega_{\rm survey}$ is the total survey area, and
$\Omega_{{\rm eff},c}$ is the effective survey area within the pixel. 
Cells with $\mathrm{f}<0.005$ are excluded to avoid poorly constrained
effective volumes near survey boundaries.

After applying the magnitude, redshift, and area cuts, the final
sample contains $N_{\rm 6dF}=53437$, $N_{\rm 2MRS}=29722$, and
$N_{\rm GAMA}=7739$ galaxies. Counting the populated cells of each
survey separately gives 882 survey cells (581 2MRS cells + 300 6dF cells + 1 GAMA cell). Of these, 297 correspond to
regions where 2MRS and 6dF occupy the same angular pixel and redshift
shell. These pairs represent the same underlying volume and are
therefore matched to a single spatial cell in the hierarchical model.
The two surveys then jointly constrain the luminosity function in that
cell while retaining their own magnitude limits and selection
functions. Removing these duplicated representations leaves
$882-297=585$ unique spatial regions in the final model.

\subsection{Luminosity Function Parameterisation}
\label{subsec:LF_param}

The luminosity function is modelled using the Schechter form \citep{Schechter1976}. 
For a single component, this is written as
\begin{align}
\phi(M,z)
&=
0.4\ln(10)\,\phi^\ast(z)\,
10^{0.4(\alpha+1)(M^\ast(z)-M)}
\nonumber \\
&\quad \times
\exp\!\left[-10^{0.4(M^\ast(z)-M)}\right],
\end{align}
where $\alpha$ is the faint-end slope, $\phi^\ast$ is the normalisation, and $M^\ast$ is the characteristic magnitude.

The redshift dependence of the luminosity function is described using the standard $P$--$Q$ parameterisation \citep{1999ApJ...518..533L,2020MNRAS.499..631V},
\begin{equation}
\phi^\ast(z) \propto 10^{0.4Pz},
\qquad
M^\ast(z) = M_0 - Qz.
\end{equation}
The parameter $P$ describes density evolution, while $Q$ describes luminosity evolution. 
In the present model, these parameters should be interpreted as effective redshift-dependent terms. 
They describe the smooth evolution of the mean galaxy population across the redshift range considered here.

For evaluating the luminosity function, the relevant quantity is the galaxy magnitude relative to the characteristic magnitude, $M-M^\ast(z)$. Separating the observational redshift dependence from the luminosity-evolution term, and using $M=m-\Delta_z$, gives
\begin{equation}
M-M^\ast(z)
=
m-\Delta_z-M_0+Qz.
\end{equation}
This can equivalently be written as $m-m^\ast(z)$, where

\begin{equation}
m^\ast(z)=M_0+\Delta_z-Qz,
\end{equation}
so that the distance- and redshift-dependent terms are absorbed into the characteristic apparent magnitude describing the observed magnitude distribution at each redshift.

A $K$-correction could in principle be included as part of the redshift-dependent term $\Delta_z$, but we do not apply galaxy-by-galaxy $K$-corrections here. Any approximately linear component of the mean $K$-correction is formally degenerate with the linear evolution term $Q$. For GAMA, where individual $K$-corrections are available, the mean correction at $z=0.065$ is only $\sim0.02$ mag, with a galaxy-to-galaxy scatter of a similar size. The mean correction would therefore correspond to a shift of only $\sim0.3$ in $Q$. Thus, including or excluding the mean $K$-correction primarily changes the fitted value and interpretation of $Q$, while the small galaxy-to-galaxy scatter has no significant effect on the inferred $J$-band luminosity density.

The luminosity function used in this work is a double-Schechter function,
\begin{equation}
\phi(M,z) =
\phi_1(M,z) + \phi_2(M,z),
\end{equation}
where each component follows the Schechter form above. 
The cosmic-mean luminosity function is parameterised by $\alpha_1$, $\alpha_2$, $\phi_1^\ast$, $\phi_2^\ast$, and the characteristic magnitude $M_1^\ast$. 
The second component is written as an offset from the first,
\begin{equation}
M_2^\ast =
M_1^\ast + \Delta M_{12}.
\end{equation}

The luminosity function is effectively defined relative to the observer-frame bandpass at each redshift. 
The luminosity density measured in each shell therefore corresponds to the luminosity density under this observer-frame treatment. 
Since the main goal is to compare luminosity-density variations between cells at fixed redshift, rather than recover a fully rest-frame luminosity function across all redshifts, this approximation is sufficient for the present analysis.

\subsection{Constraining the Luminosity Function and Luminosity Density}

The observed galaxy distribution constrains the luminosity function in two complementary ways. 
The positions of galaxies in $(m,z)$ space constrain the relative shape of the luminosity function, as described above. 
However, the normalisation parameters, $\phi^\ast_1$ and $\phi^\ast_2$, set the overall scale of the luminosity function and therefore require the total number of galaxies to be modelled as well. 
For each cell, the model computes the expected number of galaxies, $N_{\rm pred}$, from the normalising integral introduced in Equation~\ref{eq:npred}. 
This predicted number is then compared to the observed number of galaxies in that cell using a Poisson counts likelihood. 
In this way, the shape and normalisation of the luminosity function are constrained simultaneously.

The luminosity density is then computed deterministically from the same fitted luminosity function. 
For each cell, $j_c$ is evaluated through the luminosity-density integral defined in Equation~\ref{eq:luminosity_density} using the cell's LF parameters. 
Thus, both the expected number of galaxies and the luminosity density are deterministic transformations of the Schechter-function parameters during each model evaluation. 
The former constrains the LF normalisation through the observed counts, while the latter provides the cell-level luminosity-density field used in the rest of the analysis.

\subsection{Hierarchical Model}

The model infers luminosity-function parameters for each spatial cell, while also fitting the parent distribution from which those cell values are drawn. 
This is the hierarchical structure: the cell-level parameters describe local variations, while the parent distribution describes the cosmic mean and the scatter around it.

Spatial variation is introduced through the two normalisations and the characteristic magnitude,
\begin{equation}
\theta_c =
\left\{
\log_{10}\phi_{1,c}^{\ast},
\log_{10}\phi_{2,c}^{\ast},
M_{1,c}^{\ast}
\right\}.
\end{equation}
For each redshift shell, these parameters are drawn from Gaussian parent distributions,
\begin{align}
\log_{10}\phi_{1,c}^{\ast}
&\sim
\mathcal{N}\!\left(\log_{10}\phi_1^\ast,\sigma_{\phi_1}(z)\right),
\nonumber \\
\log_{10}\phi_{2,c}^{\ast}
&\sim
\mathcal{N}\!\left(\log_{10}\phi_2^\ast,\sigma_{\phi_2}(z)\right),
\nonumber \\
M_{1,c}^{\ast}
&\sim
\mathcal{N}\!\left(M_1^\ast,\sigma_{M_1}(z)\right).
\end{align}
The Gaussian form is a convenient modelling choice rather than a physical
requirement. We adopt it because the inferred cell-level distributions
of $\log_{10}\phi^\ast$ and $M^\ast$ are well described by Gaussian
distributions; other parent distributions could be used if required
by the data.
The means of these distributions define the cosmic-mean luminosity function, while the scatter terms describe the allowed cell-to-cell variation in each redshift shell. 
At the same time, these parent distributions act as priors on the individual cells, so weakly constrained regions are regularised by the full sample rather than fitted independently.

For each redshift shell, we also infer a luminosity-density scatter parameter, $\sigma_j(z)$. 
The luminosity density in each cell, $j_c$, is computed from the fitted cell-level luminosity function, while $\sigma_j(z)$ describes the intrinsic scatter of these cell luminosity densities around the shell mean.

A useful feature of this construction is that the different constraints can be combined in one posterior. 
The observed $(m,z)$ distribution constrains the shape of the luminosity function, the number of galaxies in each cell constrains the normalisation, and the distribution of $j_c$ values constrains the luminosity-density scatter. 
Since all of these pieces depend on the same model parameters, their log-probability contributions are added in the joint inference. 
The full mathematical form is given in Appendix~\ref{app:model}.

\subsection{Combining Surveys and Inference}

Surveys are combined by allowing overlapping datasets to constrain the same underlying cell parameters. 
Where 2MRS and 6dF cover the same angular region and redshift shell, they share one set of luminosity-function parameters, but retain their own magnitude limits, effective areas, and selection functions. 
The likelihood contributions from the two surveys are then evaluated simultaneously for the same physical region. 
This gives a coherent description of the underlying luminosity function conditioned on both datasets, rather than separate fits that have to be compared afterwards.

\begin{figure}
    \centering
    \includegraphics[width=1.0\linewidth]{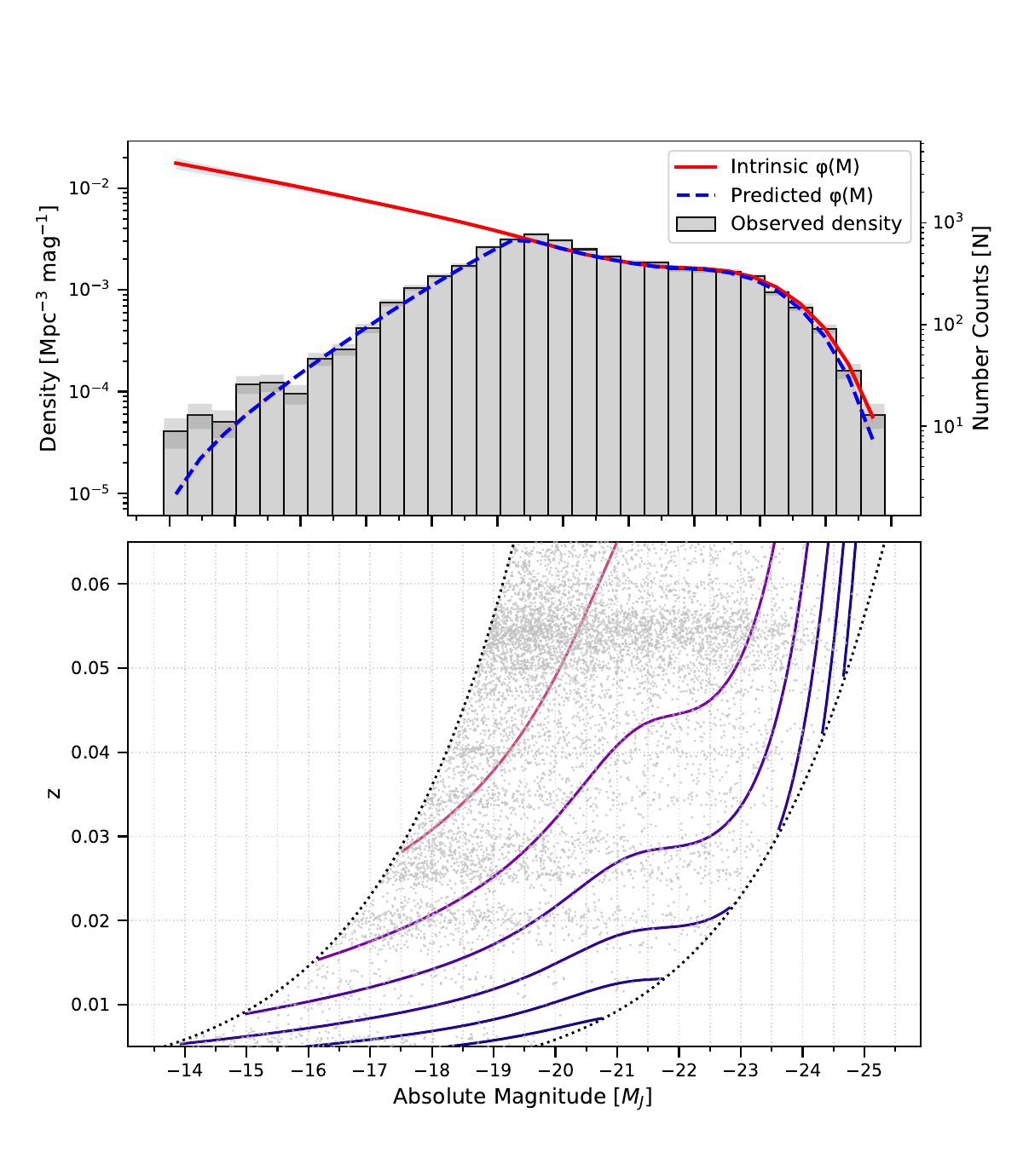}
    \caption{Luminosity-function prediction for the GAMA cell. 
    The upper panel shows the inferred luminosity function in absolute-magnitude space, with the intrinsic LF shown in red, the forward-modelled prediction after applying the GAMA selection function shown as a blue dashed curve, and the observed galaxy distribution shown as a histogram. 
    The lower panel shows the corresponding galaxy distribution in $(M,z)$ space with representative equal-density contours from the model. 
    GAMA reaches substantially fainter magnitudes than the wide-area surveys, providing the main constraint on the faint-end behaviour of the luminosity function The $(M,z)$ distribution is shown here for visualisation; the likelihood itself is evaluated in observed $(m,z)$ space.}
    \label{fig:gama_lf}
\end{figure}

\begin{figure*}
    \centering
    \includegraphics[width=\textwidth]{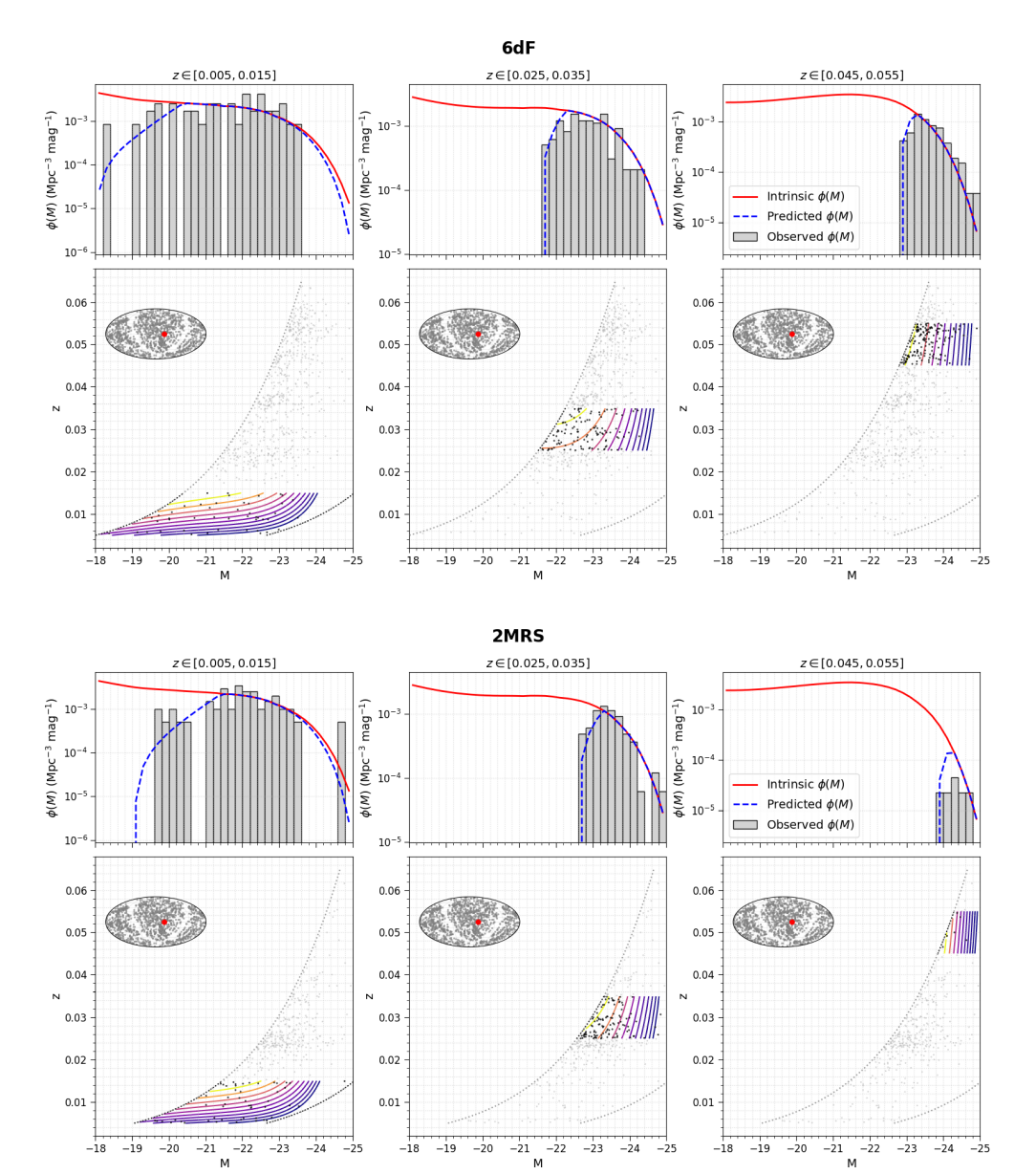}
    \caption{Example of hierarchical information sharing between spatially overlapping cells in the 6dF and 2MRS surveys. 
Each column corresponds to the same angular sky cell in a different redshift shell, while the upper and lower rows show the corresponding fits for 6dF and 2MRS respectively. 
For each survey, the upper panels show the inferred luminosity function in absolute-magnitude space. 
The red curve shows the intrinsic luminosity function, the blue dashed curve shows the forward-modelled prediction after applying the survey selection function, and the histograms show the observed galaxy distributions. 
The lower panels show the corresponding galaxy distributions in $(M,z)$ space together with representative equal-density contours predicted by the model. 
Inset sky maps indicate the location of the shared angular cell on the sky. 
Although the surveys probe different magnitude ranges and depths, the hierarchical framework constrains a common underlying luminosity function for the same physical region, allowing the surveys to converge toward a consistent solution while fitting independent datasets. The $(M,z)$ distribution is shown here for visualisation; the likelihood itself is evaluated in observed $(m,z)$ space.}
    \label{fig:cells_lf}
\end{figure*}

This sharing of information is illustrated in Figure~\ref{fig:cells_lf}. 
Although 2MRS and 6dF probe different magnitude ranges and depths, the hierarchical model fits them as distinct observations of the same underlying luminosity function in the overlapping cell. 
GAMA plays a complementary role: because it reaches substantially fainter magnitudes, it provides the main constraint on the faint-end behaviour of the luminosity function, as shown in Figure~\ref{fig:gama_lf}. 
By treating the faint-end slopes as global parameters, this information is propagated to the wider but shallower surveys.

All parameters are sampled jointly using the No-U-Turn Sampler (NUTS) \citep{2011arXiv1111.4246H} implemented in \textsc{NumPyro}. 
\textsc{NumPyro} uses \textsc{JAX} \citep{jax2018github} for automatic differentiation and just-in-time compilation, enabling efficient gradient evaluation of the full hierarchical likelihood.

\subsection{Modelling Choices}
\label{sec:model_choices}

The key modelling choices adopted in this work are summarised
here for clarity. We formulate the likelihood in observed $(m,z)$
space so that the survey limits remain fixed within each spatial cell.
This means that the likelihood normalisation needs to be evaluated
only once per cell at each model evaluation. Applying galaxy-by-galaxy
$K$-corrections or extinction corrections would make the effective
limits different for each galaxy and would therefore require a
separate integral for every object. We avoid this by treating these
effects at the cell or population level, which keeps the hierarchical
inference computationally manageable, while allowing selection effects to be applied directly to the model prediction.

We use a double-Schechter parameterisation because a single-Schechter function is usually sufficient for shallow wide-area surveys such as previous 2MASS- and 6dF-based measurements \citep{Cole2001,Jones2006}, but does not fully capture the faint-end behaviour probed by GAMA. 
GAMA luminosity-function studies show evidence for a faint-end steepening or upturn that is not captured by a single-Schechter form \citep{2012MNRAS.420.1239L,2020MNRAS.499..631V}, and a similar two-component structure is also required in GAMA stellar-mass-function measurements \citep{2012MNRAS.421..621B,2017MNRAS.470..283W}. 
The double-Schechter form is therefore the simplest parameterisation that can describe both the bright and faint galaxy populations while remaining constrained by the data. 
We allow two independent faint-end slopes and two independent normalisation components, but parameterise the second characteristic magnitude as an offset from the first through $\Delta M_{12}$, because two fully independent characteristic magnitudes introduce strong degeneracies between the two Schechter components.

Not all luminosity-function parameters are allowed to vary spatially. 
The faint-end slopes $\alpha_1$ and $\alpha_2$, the magnitude offset $\Delta M_{12}$, and the evolution parameters $P$ and $Q$ are treated as global quantities. 
The faint-end slopes are mainly constrained by the deeper GAMA data, while the shallower 2MRS and 6dF cells do not contain enough faint galaxies to constrain the faint-end behaviour independently. 
Allowing these parameters to vary freely in every cell would therefore introduce poorly constrained degeneracies with the normalisation parameters. 
Finally, $k_{\mathrm{6dF}}$ is a free parameter fitted jointly with the other model parameters, that accounts for the small residual mismatch between the 6dF random catalogue and the observed galaxy counts, bringing the effective completeness of 6dF and 2MRS onto a common footing within the joint fit.

\section{Results}
\label{sec:results}

\subsection{Luminosity Function and Evolution}

\begin{figure*}
    \centering
    \includegraphics[width=\textwidth]{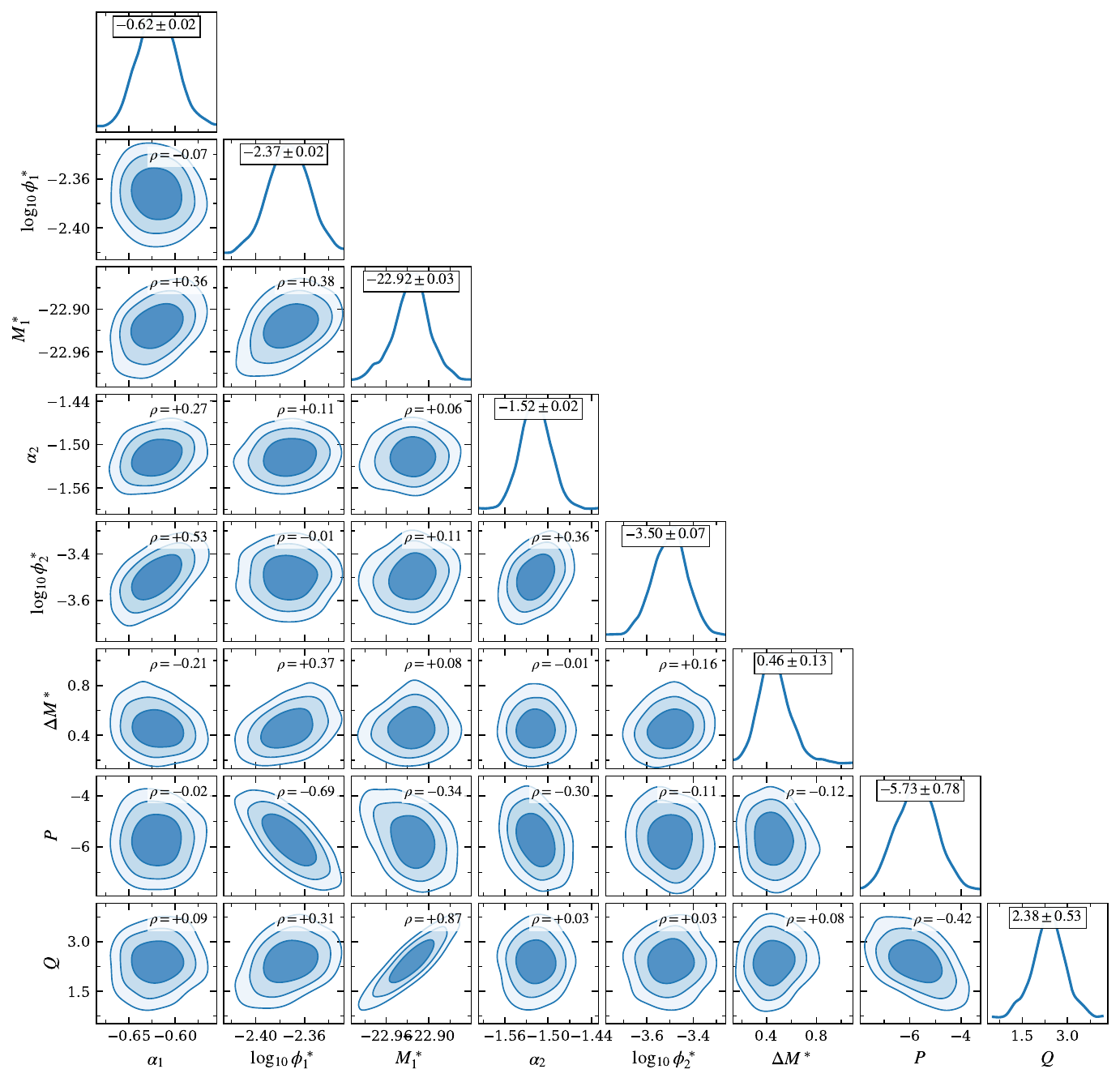}
    \caption{
    Corner plot showing the joint posterior constraints for the global luminosity-function parameters,
    $\alpha_1$, $\alpha_2$, $M_0$, $\Delta M_{12}$, $\log_{10}\phi_1$, $\log_{10}\phi_2$, $P$, and $Q$.
    The diagonal panels show the one-dimensional marginalised posterior distributions, labelled by the posterior mean and standard deviation. 
    The off-diagonal panels show the corresponding two-dimensional posterior contours, together with the linear correlation coefficient $\rho$ in each pairwise projection.}
    \label{fig:global_corner}
\end{figure*}

Table~\ref{tab:params} summarises the principal model parameters inferred by the hierarchical model, while Figure~\ref{fig:global_corner} shows the corresponding joint posterior constraints for the global luminosity-function parameters.
The model contains six parameters describing the cosmic-mean $z=0$ double--Schechter luminosity function, 
$\{\alpha_1,\alpha_2,M_0,\Delta M_{12},\log\phi_1,\log\phi_2\}$, 
together with two parameters, $P$ and $Q$, describing its smooth redshift evolution. 
The cell-level luminosity functions are described by spatially varying values of 
$\log\phi_{1,c}$, $\log\phi_{2,c}$, and $M_{1,c}^{\ast}$, whose distributions around the cosmic mean are controlled by redshift-dependent scatter hyperparameters. 
These hyperparameters are constrained by the ensemble of cells, while also acting as priors on the individual cell-level parameters. 
In total, the model jointly constrains 1792 parameters across all spatial cells and redshift shells.

\begin{table*}
\centering
\caption{
Summary of the principal model parameters.
The global parameter covariances are shown separately in Figure~\ref{fig:global_corner}.
Universal parameters describe the cosmic mean double--Schechter luminosity function,
while hyperparameters define the per--cell variations and redshift--dependent scatter.
For array parameters, each element corresponds to one spatial cell in the hierarchical model.}
\label{tab:params}
\begin{tabular}{
p{2.0cm}  
p{5.0cm}  
p{2.5cm}  
p{3.0cm}  
}
\hline
\hline
Parameter & Description & Mean $\pm1\sigma$ & Prior \\
\hline
$\alpha_1$ &
Faint--end slope of the primary Schechter component &
$-0.62\pm0.02$ &
$\mathrm{Uniform}(-3,\,2)$ \\[3pt]

$\alpha_2$ &
Faint--end slope of the secondary component &
$-1.52\pm0.02$ &
$\mathrm{Uniform}(-3,\,2)$ \\[3pt]

$M_0$ &
Characteristic magnitude of the bright component at $z{=}0$ &
$-22.92\pm0.03$ &
$\mathrm{Uniform}(-24,\,-16)$ \\[3pt]

$\Delta M_{12}$ &
Offset between bright and faint components ($M_2^*-M_1^*$) &
$0.46\pm0.13$ &
$\mathrm{Uniform}(0,\,5)$ \\[3pt]

$\log_{10}\phi_1$ &
Normalisation of the primary component [Mpc$^{-3}$] &
$-2.37\pm0.02$ &
$\mathrm{Uniform}(-5,\,0)$ \\[3pt]

$\log_{10}\phi_2$ &
Normalisation of the secondary component [Mpc$^{-3}$] &
$-3.50\pm0.07$ &
$\mathrm{Uniform}(-5,\,0)$ \\[3pt]

$P$ &
Density--evolution parameter ($\phi^*(z)\propto10^{0.4Pz}$) &
$-5.73\pm0.78$ &
$\mathrm{Normal}(0,\,10)$ \\[3pt]

$Q$ &
Luminosity--evolution parameter ($M^*(z)=M_0-Qz$) &
$2.38\pm0.53$ &
$\mathrm{Normal}(0,\,10)$ \\[3pt]

$k_{\mathrm{6dF}}$ &
Multiplicative completeness factor for 6dF &
$1.079\pm0.009$ &
$\mathrm{LogUniform}(0.01,\,10)$ \\[3pt]

$\log_{10}\phi_{1,c}$ &
Array of per--cell normalisations for the primary component; one value per spatial cell &
--- &
$\mathrm{Normal}(\log_{10}\phi_1,\,\sigma_{\phi_1,z})$ \\[3pt]

$\log_{10}\phi_{2,c}$ &
Array of per--cell normalisations for the secondary component; one value per spatial cell &
--- &
$\mathrm{Normal}(\log_{10}\phi_2,\,\sigma_{\phi_2,z})$ \\[3pt]

$M_{1,c}^{\ast}$ &
Array of per--cell characteristic magnitudes; one value per spatial cell &
--- &
$\mathrm{Normal}(M_0,\,\sigma_{M_0,z})$ \\[3pt]
\hline
\end{tabular}
\end{table*}

\begin{figure*}
    \centering
    \includegraphics[width=\textwidth]{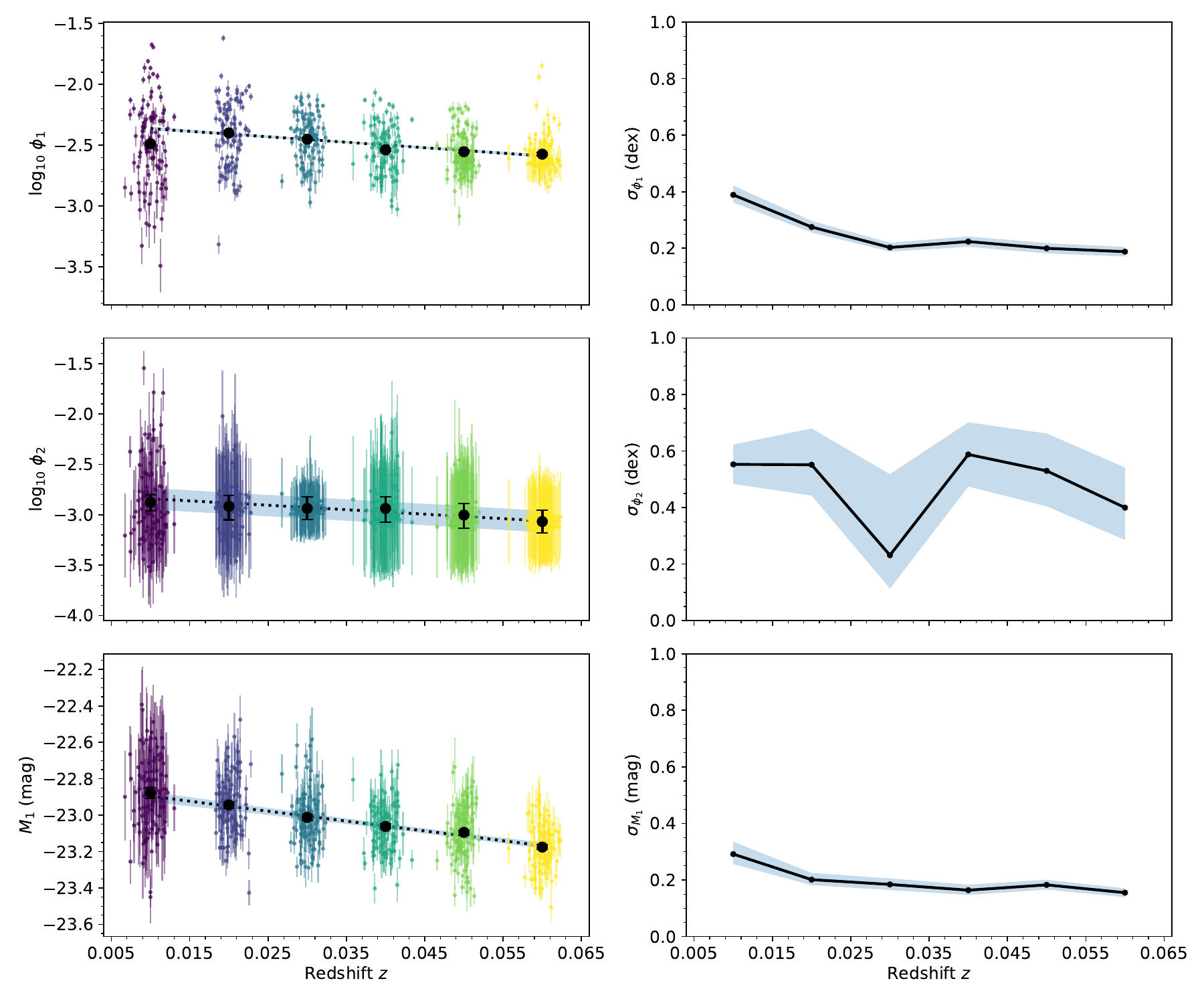}
    \caption{
        Redshift evolution of the luminosity-function parameters.
        The left panels show the cell-level posterior values of
        $\log_{10}\phi_1$, $\log_{10}\phi_2$, and $M_1^\ast$, with coloured
        points representing individual spatial cells. The points are slightly
        offset around the midpoint of each contiguous redshift shell for
        visual clarity. Black symbols show the shell means, while the solid
        black lines show the evolving cosmic mean implied by the global
        luminosity-function parameters. The right panels show the corresponding
        redshift-dependent scatter parameters $\sigma_{\phi_1}$,
        $\sigma_{\phi_2}$, and $\sigma_{M_1}$.
        }

    \label{param_z}
\end{figure*}

Figure~\ref{param_z} shows the redshift evolution of the luminosity--function parameters. 
The left panels display the cell-level posterior means in each redshift shell. 
Coloured points correspond to individual spatial cells. 
Black symbols indicate the shell mean with the $16$--$84$ percentile range. 
The solid line traces the evolving cosmic mean implied by the universal hyperparameters and the evolution parameters $P$ and $Q$.

The primary normalisation $\log\phi_1$ shows a clear redshift
trend, declining gradually with increasing redshift.
The inferred density-evolution parameter is $P=-5.73\pm0.78$,
corresponding to a modest reduction in the number density of luminous
galaxies across the fitted interval.
The lowest redshift shell lies below the cosmic-mean trend, indicating
an underdensity at $z\simeq0.01$, discussed further in
Sections~\ref{sec:j_density_evol} and
\ref{sec:significance_under_density}.
Although $P$ is well constrained by the model, its physical
interpretation is less straightforward over this short redshift
baseline. The measured trend can reflect a combination of genuine
evolution in the galaxy population and spatial density variations
across the local volume. We therefore interpret $P$ as describing the
smooth mean density trend with redshift, rather than as a measurement
of cosmological density evolution alone.

The secondary normalisation $\log\phi_2$ follows a similar redshift dependence, although the right-hand panels show that $\sigma_{\phi_2}(z)$ remains comparatively large and weakly constrained. 
This reflects the limited leverage on the faint component in individual cells.

The characteristic magnitude $M_1^\ast$ is well constrained and brightens with redshift. 
The inferred luminosity-evolution parameter is $Q=2.38\pm0.53$, corresponding to a $\sim0.14$~mag brightening across the fitted redshift range. 
The scatter $\sigma_{M_1}(z)$ remains approximately constant at $\sim0.2$~mag and is smaller than the scatter in the normalisation parameters.

The right-hand panels of Figure~\ref{param_z} show the redshift dependence of the hyperparameter scatters. 
The scatter in $\phi_1$ decreases steadily from $\sim0.45$~dex at $z\simeq0.01$ to $\sim0.2$~dex at $z\simeq0.06$. 
This is most naturally understood as a consequence of the increasing comoving volume of each angular cell at higher redshift: as larger volumes are averaged, the cell-to-cell variation in the luminosity-function normalisation decreases. 
This behaviour is consistent with the expected reduction in cosmic variance, discussed further in Section~\ref{sec:discussion}. 
The scatter in $M_1^\ast$ shows no strong redshift trend, while $\sigma_{\phi_2}$ remains weakly constrained.

\begin{figure}
    \centering
    \includegraphics[width=\linewidth]{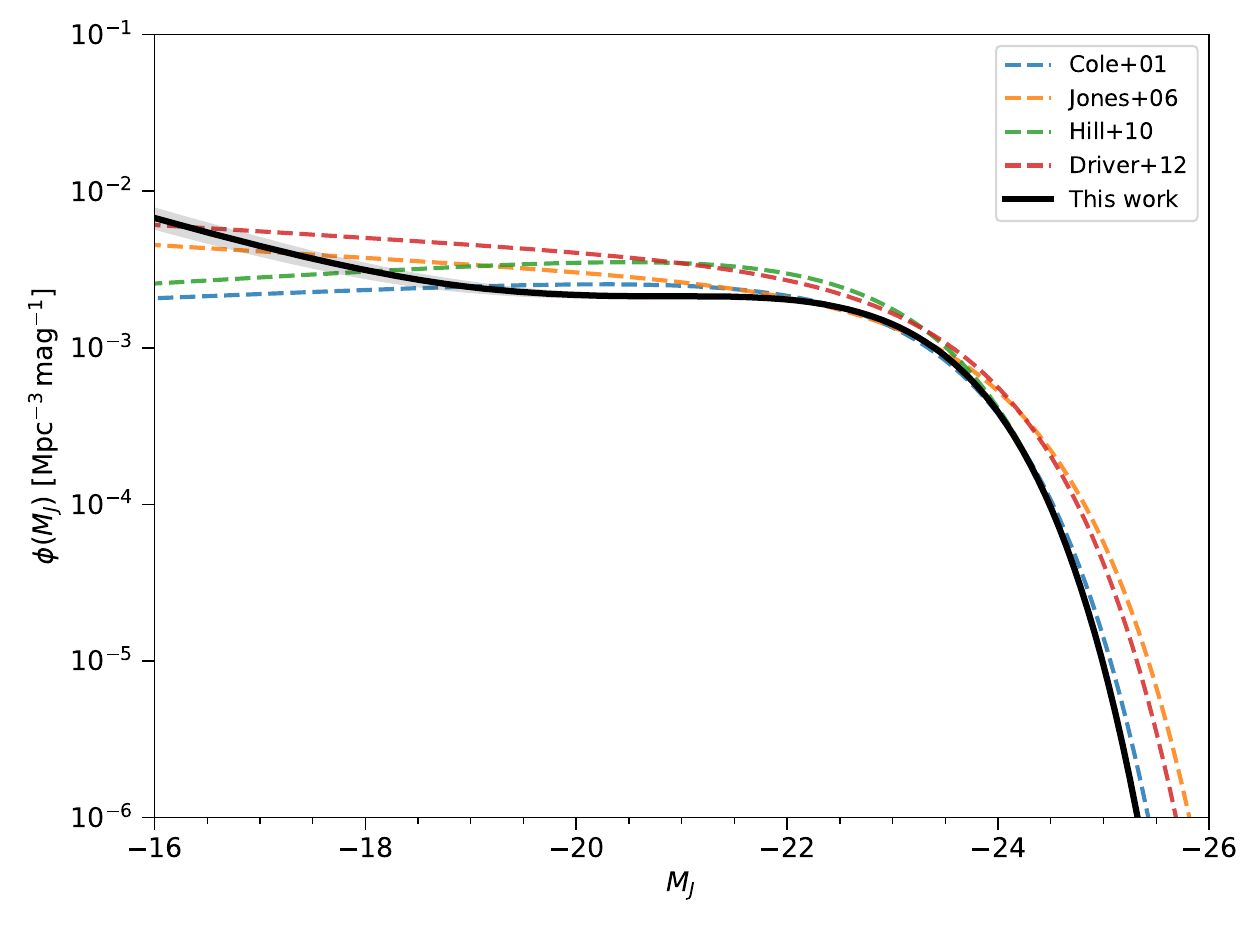}
    \caption{Comparison of the inferred $J$-band luminosity function with previous low-redshift measurements. 
    All luminosity functions have been placed on a common $H_0=70~\mathrm{km\,s^{-1}\,Mpc^{-1}}$ cosmology for comparison. 
    Measurements originally reported in the AB system, including Driver+12, have been converted to the Vega system using equation \ref{eq:ABtoVega}.}
    \label{lf_comparison}
\end{figure}

Table~\ref{tab:LF_compare} compares the recovered cosmic-mean $J$-band luminosity--function parameters with previous low-redshift measurements. Figure~\ref{lf_comparison} shows the same comparison.
The recovered luminosity function is broadly consistent with previous low-redshift measurements, although the individual Schechter parameters are not directly comparable because we use a double-Schechter form. 
The faint-end slopes $\alpha_1=-0.62\pm0.02$ and $\alpha_2=-1.52\pm0.02$ reproduce the well-known faint-end upturn seen in optical and near-infrared surveys \citep[e.g.][]{2005ApJ...631..208B}. 
The recovered mean luminosity density,
$\log_{10} j \simeq 8.45\,[L_{\odot,J}\,\mathrm{Mpc}^{-3}]$,
lies within the canonical range of previous measurements \citep{2001MNRAS.326..255C,2012MNRAS.427.3244D}.
This agreement shows that the hierarchical framework recovers a physically meaningful luminosity function while allowing for spatial variation.

The inferred luminosity density is robust to the treatment of the second characteristic magnitude in the double--Schechter model. 
We tested alternative formulations in which the two components share a common characteristic magnitude ($\Delta M_{12}=0$) or are allowed to vary independently. 
In both cases, the resulting $j_c$ values and the mean luminosity density as a function of redshift remain unchanged within uncertainties. 
Thus, while the two slopes and two normalisations are important for describing the luminosity function, the luminosity-density measurements are not strongly affected by the precise parameterisation of $M_2^\ast$.

\begin{table*}
    \centering
    \caption{Comparison of $J$-band luminosity function parameters with previous low-redshift studies.}
    \label{tab:LF_compare}
    \begin{tabular}{lccccc}
        \hline
        Parameter 
        & Cole+01 
        & Jones+06 
        & Hill+10 
        & Driver+12 
        & \textbf{This work} \\
        \hline
        $M^*_{J,1}-5\log_{10}h$     
            & $-22.36$ & $-22.85$ & $-22.20$ & $-22.68$ & \textbf{$-22.15\pm0.03$} \\
        $M^*_{J,2}-5\log_{10}h$     
            & --- & --- & --- & --- & \textbf{$-21.69\pm0.03$} \\
        $\alpha_1$     
            & \multirow{2}{*}{$-0.93$} 
            & \multirow{2}{*}{$-1.10$} 
            & \multirow{2}{*}{$-0.90$} 
            & \multirow{2}{*}{$-1.10$} 
            & \textbf{$-0.62\pm0.02$} \\
        $\alpha_2$               
            & & & & & \textbf{$-1.52\pm0.02$} \\
        $\log_{10}\phi_1\,[h^3\,{\rm Mpc}^{-3}]$
            & \multirow{2}{*}{$-1.98$} 
            & \multirow{2}{*}{$-2.15$} 
            & \multirow{2}{*}{$-1.81$} 
            & \multirow{2}{*}{$-2.01$} 
            & \textbf{$-1.91\pm0.02$} \\
        $\log_{10}\phi_2\,[h^3\,{\rm Mpc}^{-3}]$
            & & & & & \textbf{$-3.03\pm0.08$} \\
        $\log_{10}\rho_J\,[h\,L_{\odot,J}\,{\rm Mpc}^{-3}]$
            & $8.44$ & $8.48$ & $8.50$ & $8.54$ & \textbf{$8.45\pm0.05$} \\
        \hline
    \end{tabular}

    \vspace{0.5em}
    \begin{minipage}{0.98\textwidth}
    \footnotesize
    Notes. Literature measurements are single-Schechter fits, while this work uses a double-Schechter form. 
    Thus, the literature entries spanning the $\alpha_1$--$\alpha_2$ and $\phi_1$--$\phi_2$ rows correspond to the single-Schechter parameters $\alpha$ and $\phi^*$, rather than to individual double-Schechter components. 
    All magnitudes are quoted in the Vega system as $M_J-5\log_{10}h$; the Driver+12 value has been converted from AB using equation~\ref{eq:ABtoVega}. 
    The Schechter normalisations are quoted in units of $h^3\,{\rm Mpc}^{-3}$, and luminosity densities are given as $\log_{10}\rho_J$ in units of $h\,L_{\odot,J}\,{\rm Mpc}^{-3}$.
    \end{minipage}
\end{table*}

\subsection{Luminosity Density Maps}

\begin{figure*}
    \centering
    \includegraphics[
        width=\textwidth,
        trim={0 8.0cm 0 0},
        clip
    ]{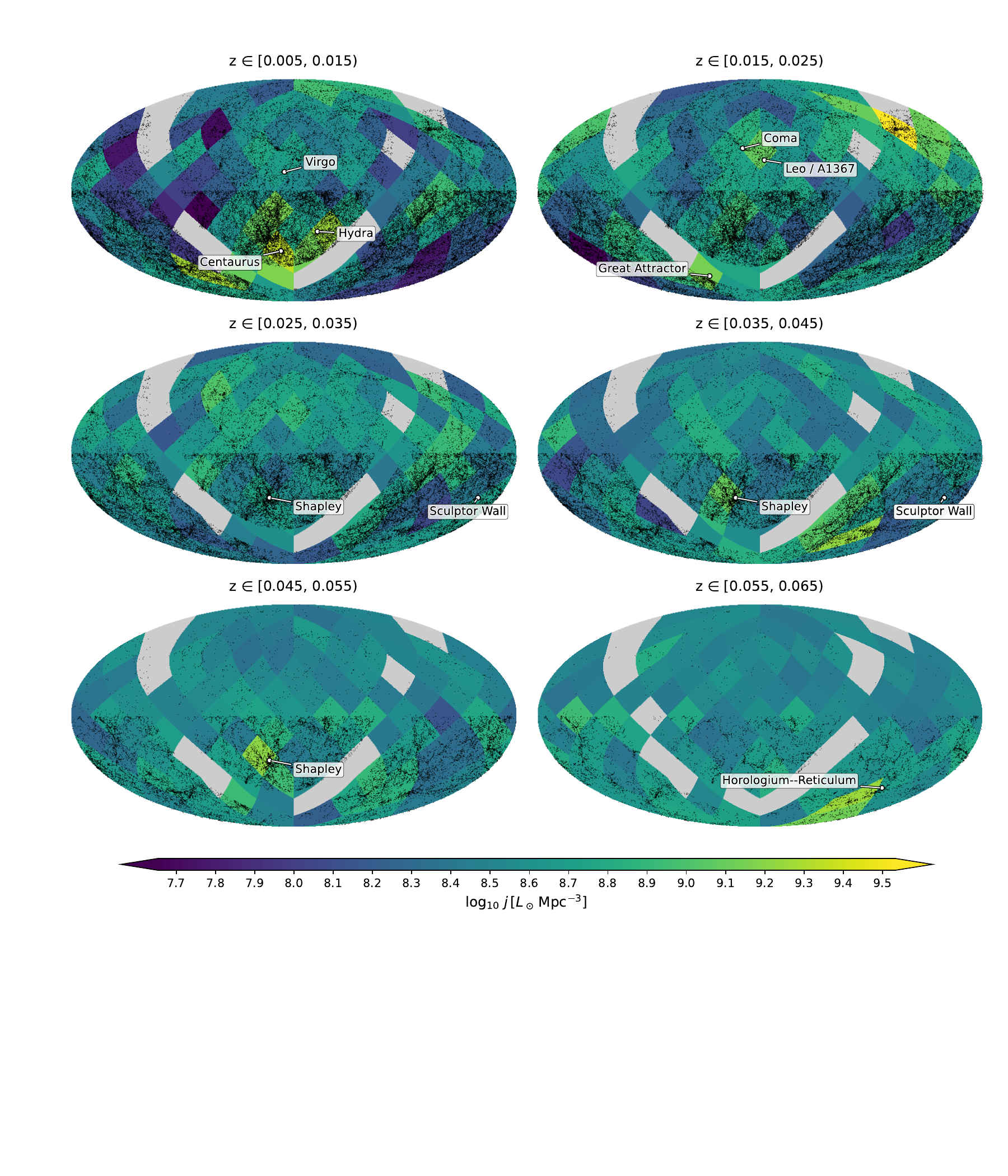}
    \caption{Luminosity-density maps in six redshift shells. 
    Colours show the posterior mean $J$-band luminosity density in each spatial cell. 
    Labels indicate prominent nearby structures.}
    \label{fig:healpix_J_shell}
\end{figure*}

Figure~\ref{fig:healpix_J_shell} shows the resulting Healpix maps across six redshift shells. 
Prominent features and large-scale structures are visible, including overdense regions associated with Shapley and Hydra--Centaurus, and are well traced by our fitted maps.

At higher redshift the distribution becomes smoother as larger volumes are sampled. 
The limited depth of 2MRS becomes apparent, particularly in the northern hemisphere where the number of galaxies per cell decreases substantially. 
In these regions, the cell-level parameters are less strongly constrained by the data and are therefore more strongly influenced by the shared parent distributions. 
This behaviour reflects the reduced information content of the data rather than an intrinsic absence of structure. 
Where the data contain sufficient galaxy counts, the likelihood can still move the cell-level parameters away from the cosmic mean; in the limit of very weak data, the inferred values naturally approach the distribution set by the hyperparameters. 
Thus, the model remains sensitive to structure where the data support it, while avoiding unstable cell-level estimates in poorly sampled regions.

We have verified that the inferred luminosity density is not driven by the sparse high-redshift sampling of 2MRS. 
Restricting the 2MRS sample to $z<0.045$, where its completeness is highest, yields luminosity-density measurements that are consistent with those obtained when including the full redshift range to $z=0.065$. 
This demonstrates that the hierarchical model effectively balances the limited 2MRS information at higher redshift with constraints from other surveys and from the global hyperparameters. 
The framework therefore allows sparsely sampled regions to still contribute to the fit, while assigning them larger uncertainties.

\subsection{Redshift Evolution of Luminosity Density}
\label{sec:j_density_evol}

\begin{figure*}
    \centering
    \includegraphics[width=\textwidth]{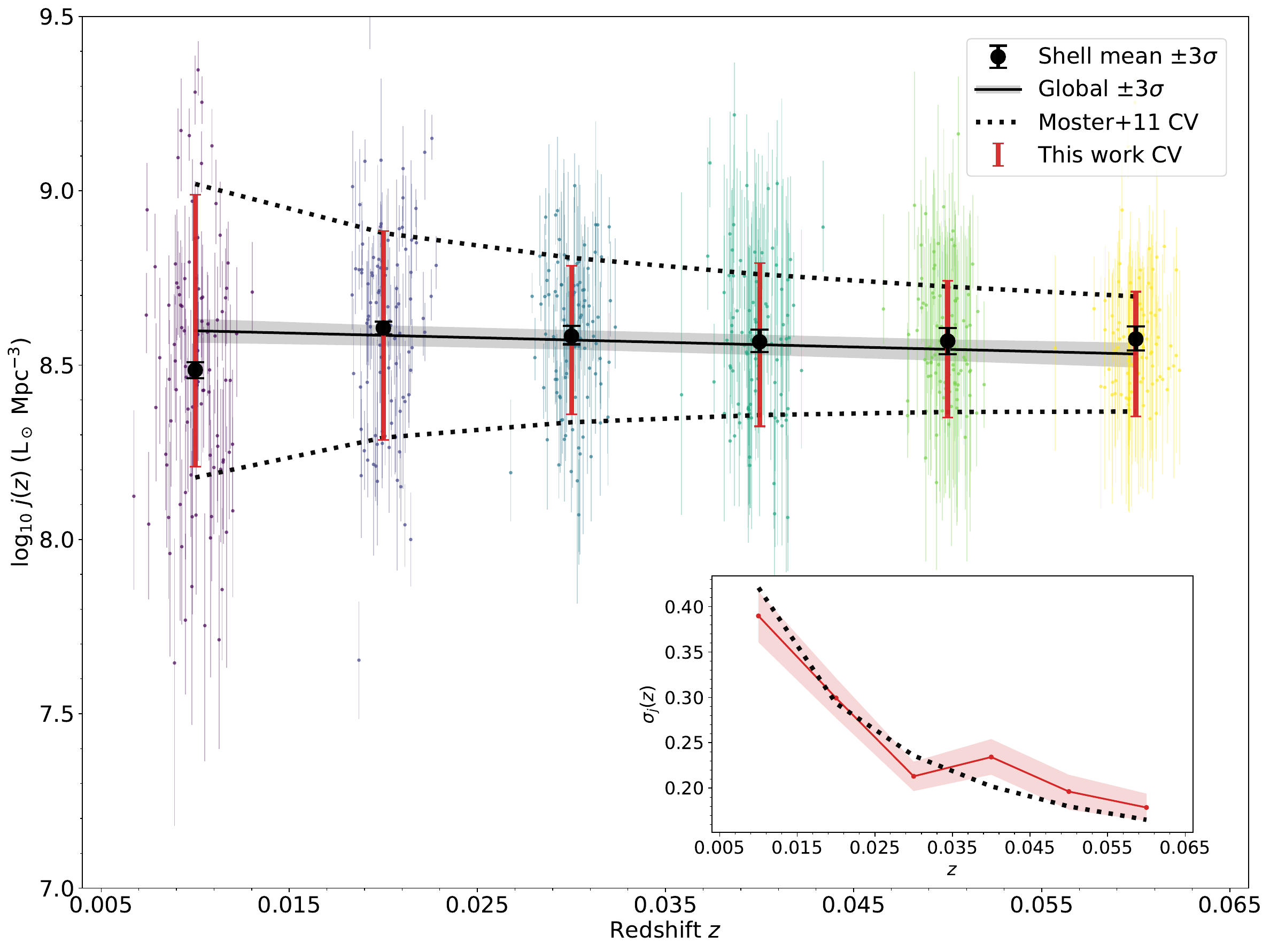}
    \caption{
        Luminosity density versus redshift for individual cells and shell
        means. The individual cell points are scattered around the centres of the contiguous redshift shells for visual clarity. The solid line shows the evolving cosmic mean. The dotted
        line shows the analytic cosmic-variance prediction of
        \citet{Moster2011}. The red error bars show the inferred
        cell-to-cell luminosity-density scatter, $\sigma_J$, within each
        redshift shell. The inset shows the redshift dependence of this
        cell-to-cell scatter compared with the theoretical expectation.}
    \label{j_z}
\end{figure*}

Figure~\ref{j_z} shows the luminosity density as a function of redshift. 
Coloured points represent the posterior luminosity densities of individual spatial cells, while black symbols show the shell-averaged luminosity density. 
The lowest redshift shell lies below the evolving cosmic mean, but the luminosity density rises toward the mean at higher redshift. 
The same figure also provides an empirical measurement of the cell-to-cell luminosity-density scatter, $\sigma_j(z)$, fitted by the hierarchical model in each redshift shell. 
The inset panel compares this inferred scatter directly with the analytic cosmic-variance prediction from \citet{Moster2011}.
The \citet{Moster2011} argument is simple, and assumes linear bias, based on abundance matching: which is one-to-one relation between halo mass and galaxy mass and uses perfect 'pencil-beam' geometry.
Despite the simplifying assumptions in the analytic estimate, the measured scatter follows the expected decline with redshift and lies within the predicted range. 
This indicates that the observed level of luminosity-density variation is consistent with fluctuations expected in a $\Lambda$CDM universe.

\subsection{Do we live in a significant underdensity?}
\label{sec:significance_under_density}

Finally, we turn to the curious result that the nearest redshift shell appears underdense relative to the smooth global model. 
Some scatter of the shell means around the evolving cosmic mean is expected, since each shell samples a finite and structured volume. 
However, the lowest-redshift shell is the largest outlier in our fitted volume, motivating a more direct estimate of its significance.

To evaluate the statistical significance of this result, we use the standard result that the uncertainty on the mean of $N$ independent samples drawn from a distribution of scatter $\sigma$ is $\sigma/\sqrt{N}$. 
Treating the spatial cells in each redshift shell as independent to first order, we compute the standardised deviation of the shell mean from the cosmic mean as
\begin{equation}
Z(z) = 
\frac{\log \bar{\rho}_j(z) - \log \rho_{j,0}(z)}
{\sigma_{\log\rho_j}(z)/\sqrt{N_{\rm cells}}} .
\end{equation}
Here $\bar{\rho}_j(z)$ is the shell-averaged luminosity density, $\rho_{J,0}(z)$ is the evolving cosmic mean predicted by the global luminosity-function model, $\sigma_{\log\rho_J}(z)$ is the inferred cell-to-cell luminosity-density scatter, and $N_{\rm cells}$ is the number of spatial cells in the shell.

\begin{figure}
    \centering
    \includegraphics[width=\linewidth]{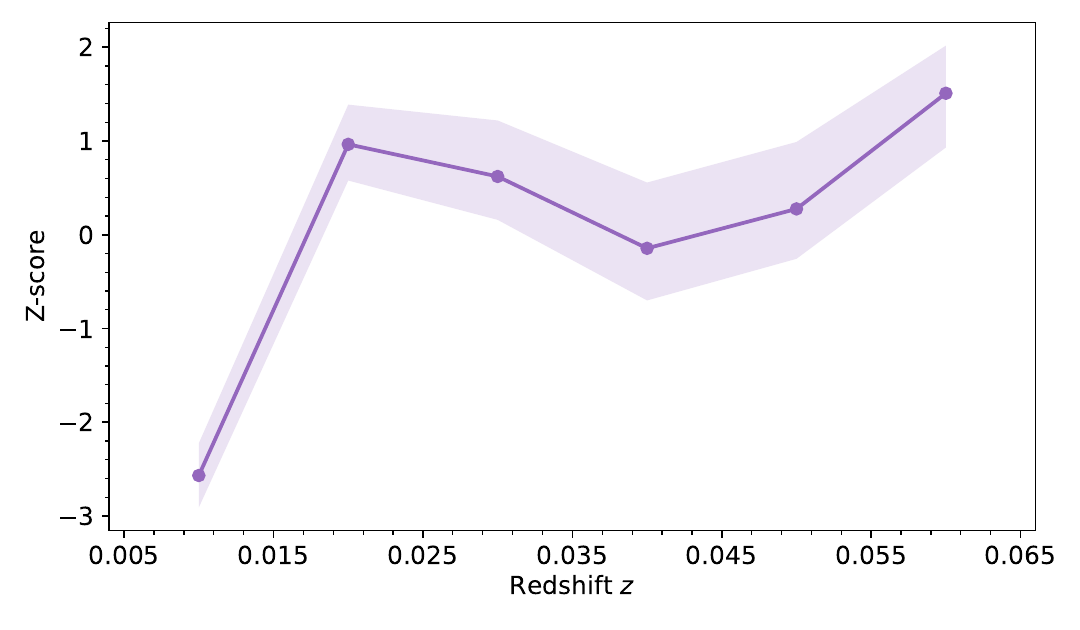}
    \caption{Standardised deviation of the shell-averaged luminosity density from the evolving cosmic mean. 
    The lowest-redshift shell is the largest outlier, with $Z\simeq-2.6$, corresponding to a one-sided significance of approximately $99.5\%$. 
    Higher-redshift shells are consistent with the cosmic mean within the inferred uncertainties.}
    \label{delta_sigmaJ_vs_z}
\end{figure}

Figure~\ref{delta_sigmaJ_vs_z} shows this quantity as a function of redshift. 
The lowest redshift shell shows the strongest deviation, with $Z\simeq-2.6$. 
This corresponds to a one-sided significance of approximately $99.5\%$, or a two-sided significance of approximately $99.1\%$. 
The underdensity occurs at $z\simeq0.01$, corresponding to a physical scale of only $\sim40$--$50$~Mpc. 
At higher redshift, the shell means scatter around the evolving cosmic mean at much lower significance.

This suggests that the deficit is a very local feature, rather than the inner part of a much larger coherent underdensity.
Instead, the result points to a very local luminosity-density deficit. 
This is not in strong tension with $\Lambda$CDM, since fluctuations of this scale can occur in finite local volumes, but its amplitude is curious and motivates further study with deeper and more uniform nearby data.

\section{Discussion and Summary}
\label{sec:discussion}

The main contribution of this work is the development and application of a hierarchical framework for mapping spatial and redshift-dependent variations in the luminosity density via the underlying LF parameters.
The model combines surveys with different depths, sky coverage, and selection functions within a single forward-modelled likelihood. 
Rather than estimating luminosity functions independently in separate sky regions, the framework jointly infers the cosmic-mean luminosity function, the cell-level deviations from that mean, and the redshift-dependent scatter in luminosity density. 
This allows the data to constrain real spatial variation where the sampling is strong, while still giving stable estimates in regions where the data are sparse.

The surveys contribute in different ways. GAMA provides the main faint-end constraint, while 2MRS and 6dF constrain the bright end, the normalisation, and the large-scale luminosity-density field.
The model therefore does not require all surveys to have the same depth or footprint; instead, their different selection functions are included directly in the likelihood.

Applied to 2MRS, 6dF, and GAMA, the model recovers a physically consistent $J$--band luminosity function. 
The inferred global luminosity-function parameters are in good agreement with previous low-redshift measurements \citep{2001MNRAS.326..255C,2006MNRAS.369...25J,2012MNRAS.427.3244D, 2010MNRAS.404.1215H}. 
The recovered mean luminosity density, $\log\rho_J\simeq8.45$, also lies within the canonical $8.3$--$8.5$ range reported in previous studies. 
This agreement is an important validation of the method. 
Although the model is fitted directly in observed $(m,z)$ space and allows the luminosity function to vary between spatial cells, it still recovers the established global luminosity-function behaviour.

The fitted evolution parameters should be interpreted as effective redshift-dependent terms within the forward model. 
We find $P=-5.73\pm0.78$ and $Q=2.38\pm0.53$, which describe the smooth evolution needed to match the observed galaxy distribution across the fitted redshift range. 
Because these terms also absorb residual redshift-dependent effects in our modelling, we do not compare their values directly to previous luminosity-function evolution measurements. 
Instead, their main role here is to provide a stable description of the mean luminosity function with redshift, against which spatial variations in luminosity density can be measured.
This interpretation is supported by repeating the analysis with $P=Q=0$. Although the fitted $z=0$ luminosity-function parameters shift, the cosmic-mean $\log_{10}J$ changes by at most $\simeq0.04$ dex across the six redshift shells, while the inferred cell-to-cell scatter, $\sigma_J$, remains consistent within $1\sigma$ in every shell. The inferred luminosity-density fluctuations are therefore insensitive to whether the evolution parameters are included.

The luminosity-density results show a clear underdensity in the lowest redshift shell, at $z\simeq0.01$. 
As shown in Figure~\ref{j_z}, the shell-averaged luminosity density in this shell lies below the evolving cosmic mean at approximately the $2.6 \sigma$ level, where the uncertainty is the posterior uncertainty from the MCMC fit. 
The shell is centred at a comoving distance of approximately $40$--$50$~Mpc, and no similar deficit is seen in the higher-redshift shells.

Beyond the first shell, the luminosity density rises back toward the cosmic mean and remains broadly stable across the remaining redshift range. 
This behaviour is important for the local-hole interpretation \citep{Keenan2013}.
Our fitted volume only extends to $z=0.065$, so we cannot fully test a proposed $\sim300$~Mpc underdensity. 
However, within the volume that we do probe, we do not see a steadily increasing luminosity density with redshift. 
Such a trend would be expected if the survey volume were sampling the inner part of a very large coherent underdensity. 
Instead, the data show a local underdensity followed by luminosity densities that are consistent with the evolving mean and with the expected level of cosmic variance. 
The fact that our inferred mean luminosity density, $\log\rho_J\simeq8.45$, is also consistent with previous measurements \citep{2001MNRAS.326..255C,2006MNRAS.369...25J,2012MNRAS.427.3244D, 2010MNRAS.404.1215H} suggests that the model is not missing a large fraction of the local light. 
We therefore interpret the lowest-redshift underdensity as a nearby fluctuation in the luminosity-density field, rather than evidence for a large-scale breakdown of homogeneity.

The fitted luminosity-density scatter provides an independent consistency check on the framework. 
In the model, $\sigma_j(z)$ is inferred directly as a free parameter by comparing the luminosity densities of individual cells to the shell mean. 
The recovered scatter decreases from $\sim0.4$~dex at $z\simeq0.01$ to $\sim0.2$~dex at $z\simeq0.06$, reflecting the larger comoving volume sampled by each angular cell at higher redshift. 
This trend agrees well with the analytic cosmic-variance prediction of \citet{Moster2011}, shown by the dotted curve in Figure~\ref{j_z}. 
Given that the Moster et al.\ prediction is based on linear biasing, abundance matching, and a simplified survey geometry, this agreement is encouraging. 
It shows that the spatial variation recovered by the hierarchical model is consistent with the level of luminosity-density fluctuations expected in a standard $\Lambda$CDM universe.

This agreement is also important because the cosmic-variance prediction is not imposed on the cell luminosity densities. 
The model fits $\sigma_j(z)$ from the data, while the Moster et al.\ curve is used only as an external comparison. 
The fact that the inferred scatter follows the expected redshift dependence therefore provides a useful validation of the model. 
If the framework were dominated by selection-function errors, unstable luminosity-function fits, or survey-boundary artefacts, we would not expect the recovered luminosity-density scatter to follow the expected cosmic-variance scaling so closely.

The hierarchical framework also handles the data-poor parts of the current surveys in a natural way. 
At higher redshift, and especially in regions where only the shallower surveys contribute, the number of galaxies per cell becomes small. 
In these cases, the cell-level parameters are only weakly constrained by the data, and the posterior is therefore driven more strongly by the shared hyperparameters. 
This should not be interpreted as forcing homogeneity into the result. 
Rather, it reflects the limited information available in those cells. 
Where the data contain enough galaxies, the cell parameters remain free to move away from the mean. 
The practical advantage is that the model can make use of all available cells, including sparsely sampled ones, while fully propagating their larger uncertainties through the inference. 
There is therefore no need to discard low-count cells by hand, or to replace them with unstable independent estimates.

The same structure also allows the model to combine information across surveys. 
In overlapping regions, the luminosity-function parameters are constrained jointly by the surveys that occupy the same part of the sky. 
This is why the shallow wide-area surveys can benefit from the faint-end information provided by GAMA, while GAMA benefits from the wider-area constraints on the bright end and normalisation. 
The result is not a simple average of the surveys, but a joint fit in which each survey contributes according to its depth, footprint, and selection limits.

The inclusion of the $k_{\mathrm{6dF}}$ factor illustrates how the framework can handle residual survey-calibration uncertainty. 
The 6dF random catalogue describes the relative angular completeness of 6dF across the sky, but the overall completeness normalisation is less certain. 
By contrast, 2MRS is shallower but highly complete, and therefore provides a stronger constraint on the absolute number-density scale. 
We therefore introduce a multiplicative factor, $k_{\mathrm{6dF}}$, which calibrates the effective 6dF completeness by requiring consistency between the overlapping 6dF and 2MRS constraints. 
This allows the model to retain the main strengths of both surveys: 6dF contributes its deeper magnitude coverage and stronger leverage on the luminosity-function shape, while 2MRS anchors the overall normalisation. 
At the same time, the calibration uncertainty is inferred within the joint model rather than imposed by hand, reducing the risk that residual completeness mismatches propagate into the luminosity-density field.

In total, the model fits 1792 parameters across all spatial cells and redshift shells. 
Despite this high dimensionality, the inference remains stable and computationally tractable. 
This is made possible by evaluating the likelihood with \textsc{NumPyro} and \textsc{JAX}, which allow automatic differentiation and efficient sampling with NUTS. 
The computational design is not simply a technical detail: it is what makes it possible to fit a hierarchical luminosity-function model with many spatial cells, multiple surveys, and selection-function normalisation within a single Bayesian framework.

The present analysis is therefore best viewed as both a measurement and a demonstration of the method. 
With the current data, the model recovers a consistent low-redshift luminosity function, detects a nearby luminosity-density underdensity, and measures a redshift-dependent scatter consistent with standard cosmic variance. 
At the same time, the strength of the present test is limited by the available data, especially the depth and geometry of current spectroscopic redshift surveys. 
2MRS becomes sparse beyond $z\simeq0.03$--$0.04$ \citep{2012ApJS..199...26H}, 6dF has a median redshift of only $z\simeq0.05$ \citep{2004MNRAS.355..747J,2009MNRAS.399..683J}, and GAMA covers only a small sky area \citep{2015MNRAS.452.2087L,2011MNRAS.413..971D}. 
The fitted range, $0.005<z<0.065$, is therefore not large enough to provide a definitive test of structures extending to several hundred Mpc. 
The main limitation is the depth and geometry of the available data, not the structure of the model.

This is where the framework becomes especially relevant for upcoming surveys. 
Future datasets such as 4HS \citep{2023Msngr.190...46T}, DESI \citep{2016arXiv161100036D}, Euclid \citep{2025A&A...697A...1E}, LSST \citep{2019ApJ...873..111I}, SPHEREx \citep{2014arXiv1412.4872D}, and SKA \citep{2015arXiv150104076M} will provide much deeper and wider coverage than the current low-redshift samples. 
In particular, extending the same analysis to $z\sim0.15$ would allow the model to sample the full radial scale of proposed $\sim300$~Mpc local-underdensity scenarios, rather than only the nearest part of the volume. 
This would make it possible to test whether the luminosity-density field continues to rise with redshift, as expected for a large coherent local hole, or instead converges to the cosmic mean on smaller scales.

The same future datasets will also make the natural extensions of the framework more important. 
A major next step is to include photometric uncertainties directly in the forward model. 
In the present analysis, observed magnitudes are treated as exact once the survey selection limits and foreground-extinction corrections have been applied. 
This is a useful first approximation, but it neglects the scattering of galaxies across magnitude limits and across the luminosity function itself \citep{1913MNRAS..73..359E}. 
Because the model is already written as a forward model in apparent-magnitude and redshift space, photometric errors could be included by convolving the predicted $(m,z)$ distribution with the magnitude-error model for each survey. 
This would allow measurement uncertainties, survey selection, and luminosity-function inference to be treated within the same likelihood.
An alternative approach would be to combine surveys through a joint
maximum-likelihood treatment of the luminosity function and density
fluctuations \citep{cole2011, Loveday2015_LF}, without introducing spatial
cells. For the low-redshift $J$-band analysis considered here, the
global $K$-correction approximation is sufficient, while future
applications at higher redshift or in bluer bands will likely require
individual SED-dependent $K$-corrections.

Another important extension is to combine the luminosity-density model with independent distance and velocity information. 
Fundamental Plane \citep{1987ApJ...313...42D, 1987ApJ...313...59D} and Tully--Fisher \citep{1977A&A....54..661T} measurements provide constraints on peculiar velocities and bulk flows that are directly related to the same underlying density field \citep{1995ApJ...449..446Z, 2013AJ....146...86T, 2014MNRAS.445.2677S}. 
In principle, these data can be incorporated by adding their likelihood contributions to the same hierarchical model. 
This would move the framework beyond fitting the luminosity function alone, toward a joint model of the galaxy distribution, luminosity-density field, and velocity field. 
In this broader form, the framework could also be used to compare mocks with observations, validate survey selection functions, test cross-survey calibration, and map the stellar luminosity-density field across much larger cosmic volumes.

The method developed here is designed for exactly this situation: combining heterogeneous surveys, preserving their individual selection functions, and fitting the luminosity-density field in a common hierarchical framework. 
With the next generation of wide and deep surveys, and with extensions that include photometric errors and independent distance information, this approach provides a practical route to testing local structure and mapping the luminosity-density field with far greater precision.

\section{acknowledgements}
    We thank Chris Blake for his very useful discussion and comments on the paper. 
    We gratefully acknowledge Alice Serene and Astronomy Data and Computing Services (ADACS) for support in developing \texttt{numpyro\_schechter} package \citep{numpyro_schechter2025} which provides \textsc{NumPyro}-compatible Schechter-function and normalisation tools for Bayesian luminosity-function modelling, and is available open source.

\section{Data Availability}
The data underlying this article will be shared on reasonable request to the corresponding author.

\bibliographystyle{mnras}
\bibliography{references}

\appendix

\section{Hierarchical Model}
\label{app:model}

Our model defines a smooth, continuous two-dimensional function for the observed galaxy distribution in apparent-magnitude and redshift data space, $(m,z)$. 
The intrinsic distribution is described by a redshift-dependent double-Schechter luminosity function (LF) in absolute magnitude $M$,
\begin{equation}
\phi(M,z)
=
\phi_1(M,z)+\phi_2(M,z),
\end{equation}
where each component is
\begin{align}
\phi_i(M,z)
&=
0.4\ln(10)\,
\phi_i^\ast(z)\,
10^{0.4(\alpha_i+1)\left[M_i^{\ast}(z)-M\right]}
\nonumber \\
&\quad \times
\exp\left[
-10^{0.4\left[M_i^{\ast}(z)-M\right]}
\right].
\end{align}
Here $\alpha_1$ and $\alpha_2$ are the two faint-end slopes, $\phi_1^\ast$ and $\phi_2^\ast$ are the two normalisations, and $M_1^\ast$ and $M_2^\ast$ are the characteristic magnitudes of the two Schechter components that we use to describe the LF.

The LF is defined in absolute magnitude, while the data are measured in apparent magnitude. 
In the standard notation, these are related by
\begin{equation}
m - M = DM_{\rm cos}(z) + K(z) + E(z),
\end{equation}
where $DM_{\rm cos}(z)$ is the cosmological distance modulus, $K(z)$ is the $K$-correction, and $E(z)$ represents evolutionary corrections. 
The cosmological distance modulus is defined using the luminosity distance,
\begin{equation}
DM_{\rm cos}(z)
=
5\log_{10}
\left(
\frac{D_L(z)}{10\,{\rm pc}}
\right),
\end{equation}
with
\begin{equation}
D_L(z) = (1+z)D_C(z).
\end{equation}
This gives
\begin{equation}
DM_{\rm cos}(z)
=
5\log_{10}
\left(
\frac{D_C(z)}{10\,{\rm pc}}
\right)
+
5\log_{10}(1+z).
\end{equation}

In this work we separate the distance conversion from the redshift evolution of the luminosity function. 
For the mapping into $(m,z)$ data space, we adopt the baseline bandpass-stretching correction
\begin{equation}
K_0(z) = -2.5\log_{10}(1+z).
\end{equation}
This corresponds exactly to the correction for a flat-spectrum source.

We therefore define the effective magnitude-distance term
\begin{equation}
\mathcal{D}(z)
\equiv
DM_{\rm cos}(z) + K_0(z),
\end{equation}
or equivalently
\begin{equation}
\mathcal{D}(z)
=
5\log_{10}
\left(
\frac{D_C(z)}{10\,{\rm pc}}
\right)
+
2.5\log_{10}(1+z).
\end{equation}
The absolute magnitude used to evaluate the LF is then
\begin{equation}
M(m,z) = m - \mathcal{D}(z).
\end{equation}
The term $\mathcal{D}(z)$ is therefore only the magnitude conversion used to evaluate the LF in $(m,z)$ data space. 
It should not be confused with the redshift evolution of the LF parameters, which is described separately below.

We do not apply explicit galaxy-by-galaxy $K$-corrections in the
present model. The advantage of this treatment is that the survey
limits remain fixed within each spatial cell, so the normalisation
requires only one numerical integral per cell at each MCMC step.
Applying individual $K$-corrections would make the effective limits
object-dependent and would instead require a separate integral for
each galaxy, substantially increasing the computational cost of the
hierarchical inference.

The redshift dependence of the LF is described through density and luminosity evolution,
\begin{equation}
\phi^{*}_{i}(z)
=
\phi^{*}_{i,0}\,10^{0.4Pz},
\end{equation}
and
\begin{equation}
M_i^{*}(z)
=
M_0 + \Delta M_i - Qz,
\end{equation}
where $P$ and $Q$ are global evolution parameters. 
For the two Schechter components, we set
\begin{equation}
\Delta M_1 = 0,
\qquad
\Delta M_2 = \Delta M_{12},
\end{equation}
so that
\begin{equation}
M_2^\ast(z) = M_1^\ast(z) + \Delta M_{12}.
\end{equation}
Thus, the magnitude conversion into $(m,z)$ data space is handled through $\mathcal{D}(z)$, while the smooth redshift evolution of the LF is handled through $P$ and $Q$.

With these definitions, the LF is evaluated in $(m,z)$ data space as
\begin{equation}
\phi(M,z)
\rightarrow
\phi\left(m-\mathcal{D}(z),z\right).
\end{equation}
For a cell $c$ with solid angle $\Omega_c$, the model intensity in apparent-magnitude and redshift space is
\begin{equation}
\lambda_c(m,z)
=
\phi_c\left(m-\mathcal{D}(z),z\right)
\frac{dV_c}{dz},
\end{equation}
where
\begin{equation}
\frac{dV_c}{dz}
=
\Omega_c
\frac{cD_C^2(z)}{H_0E(z)},
\end{equation}
with
\begin{equation}
E(z)
=
\sqrt{\Omega_m(1+z)^3+\Omega_{\Lambda}}.
\end{equation}

Survey selection is applied directly in $(m,z)$ data space through fixed limits in apparent magnitude and redshift. 
For each original survey cell $c$, the expected number of galaxies is
\begin{equation}
\mu_c
=
\int_{z_{\min,c}}^{z_{\max,c}}
\frac{dV_c}{dz}
\left[
\int_{m_{\rm bright,c}}^{m_{\rm faint,c}}
\phi_c\!\left(m-\mathcal{D}(z),z\right)\,dm
\right] dz .
\end{equation}
This is the same selection integral that would be evaluated in absolute-magnitude space, written after the change of variables $M=m-\mathcal{D}(z)$. 
Here the apparent-magnitude limits, $m_{\rm bright,c}$ and $m_{\rm faint,c}$, are fixed by the survey selection. 
The redshift dependence enters through the distance term $\mathcal{D}(z)$, the comoving volume element, and the LF evolution parameters.

At fixed redshift, the inner magnitude integral has the standard Schechter incomplete-gamma form. 
For each Schechter component, we write
\begin{equation}
I_{i,c}(z)
=
\int_{m_{\rm bright,c}}^{m_{\rm faint,c}}
\phi_{i,c}\!\left(m-\mathcal{D}(z),z\right)\,dm .
\end{equation}
Using the variable
\begin{equation}
x_i(M,z)
=
10^{0.4\left[M_i^\ast(z)-M\right]},
\end{equation}
this integral can be written in terms of the upper incomplete gamma function as
\begin{equation}
I_{i,c}(z)
=
\phi_{i,c}^{\ast}(z)
\left[
\Gamma\!\left(\alpha_i+1,x_{i,{\rm faint}}(z)\right)
-
\Gamma\!\left(\alpha_i+1,x_{i,{\rm bright}}(z)\right)
\right],
\end{equation}
where
\begin{equation}
x_{i,{\rm bright}}(z)
=
10^{0.4\left[M_i^\ast(z)-\left(m_{\rm bright,c}-\mathcal{D}(z)\right)\right]},
\end{equation}
and
\begin{equation}
x_{i,{\rm faint}}(z)
=
10^{0.4\left[M_i^\ast(z)-\left(m_{\rm faint,c}-\mathcal{D}(z)\right)\right]}.
\end{equation}
For faint-end slopes with $\alpha_i<-1$, the shape parameter $\alpha_i+1$ is negative. 
We evaluate the upper incomplete gamma function using the recurrence relation
\begin{equation}
\Gamma(s,x)
=
\frac{\Gamma(s+1,x)-x^s e^{-x}}{s},
\end{equation}
applied recursively until the shifted shape parameter is in the stable positive range. 
In practice, this calculation is evaluated using the \textsc{JAX}-compatible implementation in the open-source \texttt{numpyro\_schechter} package\footnote{\url{https://github.com/alserene/numpyro_schechter}}.

The double-Schechter magnitude integral is then
\begin{equation}
I_c(z)=I_{1,c}(z)+I_{2,c}(z),
\end{equation}
and the expected number of galaxies becomes
\begin{equation}
\mu_c
=
\int_{z_{\min,c}}^{z_{\max,c}}
\frac{dV_c}{dz}\,
I_c(z)\,dz .
\end{equation}

The remaining redshift integral is evaluated numerically using
Gauss--Legendre quadrature implemented in \textsc{JAX}. This preserves
automatic differentiation through the full expected-count calculation,
which is required for NUTS-based inference. The survey magnitude limits
remain fixed in observed apparent magnitude, while the redshift
dependence is evaluated through the quadrature nodes. As a result, the
normalisation is computed once at the cell level at each model
evaluation rather than separately for every galaxy.

For the 6dF counts likelihood, we include a multiplicative completeness factor $k_{\rm 6dF}$. 
This factor rescales the expected number of galaxies entering the Poisson counts term,
\begin{equation}
\mu^{\rm cnt}_c
=
\begin{cases}
k_{\rm 6dF}\,\mu_c, & \text{for 6dF cells},\\
\mu_c, & \text{otherwise}.
\end{cases}
\end{equation}
The factor is used to account for residual uncertainty in the overall 6dF completeness relative to 2MRS. 
The galaxy-level position likelihood remains normalised using the unscaled expected counts $\mu_c$, while the Poisson counts likelihood uses $\mu^{\rm cnt}_c$.

Spatial variation is introduced through cell-specific LF parameters. 
Each unique spatial-redshift cell $s$ is assigned
\begin{equation}
\theta_s
=
\{
\log_{10}\phi_{1,s},
\log_{10}\phi_{2,s},
M_{1,s}
\}.
\end{equation}
These are drawn from global parent distributions,
\begin{align}
\log_{10}\phi_{1,s}
&\sim
\mathcal{N}
\left(
\log_{10}\phi_1,\,
\sigma_{\phi_1}(z_s)
\right),
\\
\log_{10}\phi_{2,s}
&\sim
\mathcal{N}
\left(
\log_{10}\phi_2,\,
\sigma_{\phi_2}(z_s)
\right),
\\
M_{1,s}
&\sim
\mathcal{N}
\left(
M_0,\,
\sigma_{M_1}(z_s)
\right).
\end{align}
The scatter parameters $\sigma_{\phi_1}(z)$, $\sigma_{\phi_2}(z)$,
and $\sigma_{M_1}(z)$ are fitted independently in each redshift shell.
They are treated as free hyperparameters with independent uniform
priors over $10^{-3}<\sigma<2$. The luminosity-density scatter
$\sigma_J(z)$ is similarly fitted in each shell with a uniform prior
over $10^{-3}<\sigma_J<1$.
The faint-end slopes $\alpha_1$ and $\alpha_2$, the offset
$\Delta M_{12}$, and the evolution parameters $P$ and $Q$ are global
parameters.

The model contains two data-likelihood terms and one additional hierarchical term for the luminosity-density scatter.
First, the galaxy-level likelihood uses the observed positions of all galaxies in $(m,z)$ space. 
For galaxy $j$, belonging to cell $c_j$, the model intensity is $\lambda_{c_j}(m_j,z_j)$. 
All galaxies in the same cell share the same cell-level normalisation $\mu_c$, and the full galaxy distribution is normalised over the total expected number of galaxies across all cells. 
The normalised galaxy-level contribution is
\begin{equation}
\ln\mathcal{L}_{\rm gal}
=
\sum_j
\ln \lambda_{c_j}(m_j,z_j)
-
N_{\rm gal}
\ln
\left(
\sum_c \mu_c
\right).
\end{equation}
The first term evaluates the forward model at the observed location of each galaxy, while the second term normalises the galaxy distribution over the full survey window.

Second, the observed number of galaxies in each original cell is constrained using a Poisson likelihood,
\begin{equation}
\ln\mathcal{L}_{\rm cnt}
=
\sum_c
\ln P
\left(
N_c \mid \mu^{\rm cnt}_c
\right),
\end{equation}
where $N_c$ is the observed number of galaxies in cell $c$.

Third, the luminosity density in each unique spatial-redshift cell is computed deterministically from the fitted cell-level LF. 
At the midpoint of the redshift shell containing cell $s$, this is
\begin{equation}
J_s
=
\int
L(M)\,
\phi_s(M,z_s)\,
dM,
\end{equation}
with
\begin{equation}
L(M)
=
10^{-0.4(M-M_{\odot,J})}.
\end{equation}
The corresponding cosmic-mean luminosity density in each shell, $\bar{J}_z$, is computed from the global LF parameters at the same shell midpoint.

To quantify luminosity-density fluctuations, the cell-level luminosity densities are modelled as Gaussian draws around the cosmic mean in log space,
\begin{equation}
\ln\mathcal{L}_{J}
=
\sum_s
\ln
\mathcal{N}
\left(
\log_{10}J_s
\mid
\log_{10}\bar{J}_{z_s},
\sigma_J(z_s)
\right).
\end{equation}
Here $\sigma_j(z)$ is a free hyperparameter in each redshift shell and describes the intrinsic cell-to-cell scatter in luminosity density.

The total log-likelihood is the sum of the three contributions,
\begin{equation}
\ln\mathcal{L}_{\rm tot}
=
\ln\mathcal{L}_{\rm gal}
+
\ln\mathcal{L}_{\rm cnt}
+
\ln\mathcal{L}_{J}.
\end{equation}
Posterior sampling is performed using the No-U-Turn Sampler (NUTS) implemented in \textsc{NumPyro}. 
All components of the model, including the numerical integrals, are differentiable under \textsc{JAX}, enabling efficient inference of the high-dimensional hierarchical posterior.

\section{Galactic extinction correction}
\label{app:dust}

\begin{figure*}
    \centering
    \includegraphics[width=\linewidth]{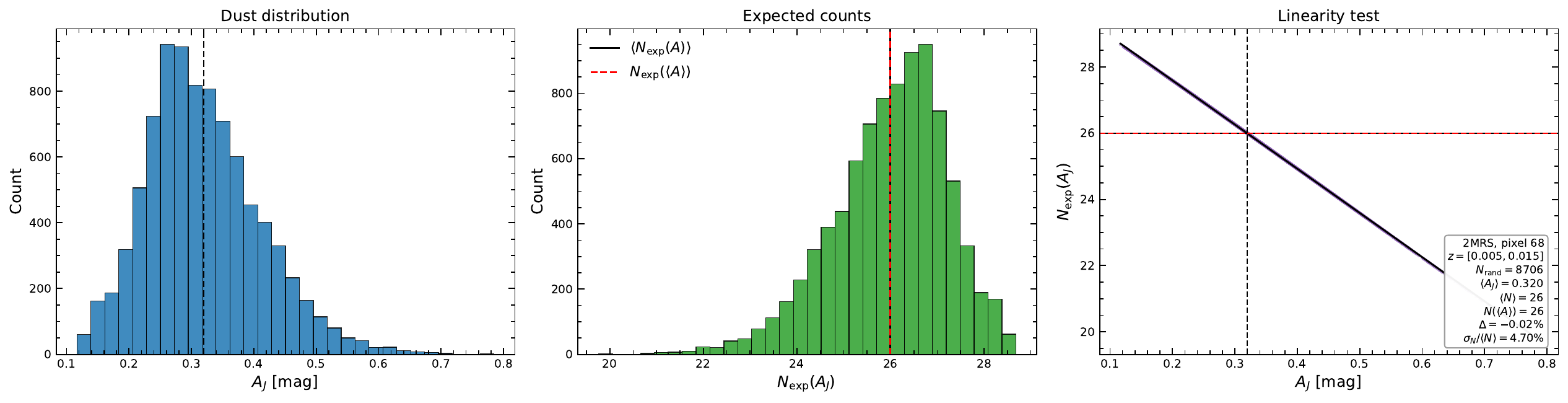}
    \caption{
    Dust variation within a representative cell and its impact on the expected number counts.
    \textit{Left:} distribution of $A_J$ values from the random catalogue.
    \textit{Middle:} distribution of $N_{\mathrm{exp}}(A_J)$ computed assuming uniform extinction.
    \textit{Right:} relation between $A_J$ and $N_{\mathrm{exp}}$, showing an approximately linear dependence.
    }
    \label{fig:dust_linearity}
\end{figure*}

Observed magnitudes are affected by Galactic dust, which reduces the observed flux of galaxies. 
For a galaxy with extinction $A_J$, the absolute magnitude corresponding to an observed apparent magnitude $m$ is
\begin{equation}
M = m - \mathcal{D}(z) - A_J,
\end{equation}
where $\mathcal{D}(z)$ is the magnitude-distance term defined in Appendix~\ref{app:model}. 
Thus, for fixed observed survey limits, Galactic extinction changes the absolute-magnitude range over which the luminosity function is sampled.

In principle, $A_J$ varies from galaxy to galaxy, and the LF argument in the normalisation integral should therefore include the local extinction at each position. 
In practice, the variation within a cell is small, so we approximate the extinction by its mean value over the cell, $\langle A_J\rangle_c$. 
The expected number of galaxies in cell $c$ is then evaluated as
\begin{equation}
\mu_c
=
\int_{z_{\min,c}}^{z_{\max,c}}
\frac{dV_c}{dz}
\left[
\int_{m_{\rm bright,c}}^{m_{\rm faint,c}}
\phi_c\!\left(m-\mathcal{D}(z)-\langle A_J\rangle_c,z\right)\,dm
\right] dz .
\end{equation}
Here the observed apparent-magnitude limits, $m_{\rm bright,c}$ and $m_{\rm faint,c}$, remain fixed by the survey selection. 
Equivalently, after the change of variables $m_0=m-\langle A_J\rangle_c$, this can be implemented as a shift of the effective apparent-magnitude limits used in the normalisation integral. 
This ensures that extinction is consistently included in both the expected counts $\mu_c$ and the normalisation term in the likelihood.

We estimate the extinction using the Schlegel dust maps \citep{Schlegel1998}. 
For each cell, $A_J$ is evaluated at the positions of the random catalogue, and the mean value is computed as
\begin{equation}
\langle A_J \rangle_c = \frac{1}{N_{\mathrm{rand},c}} \sum_i A_{J,i}.
\end{equation}
Using the randoms ensures that the estimate follows the survey footprint and mask without risk of bias towards areas of lower extinction.

Within a cell, extinction is not uniform. 
To check whether using a single value is sufficient, we compare
\begin{equation}
\langle N_{\mathrm{exp}}(A_J) \rangle
\quad \text{and} \quad
N_{\mathrm{exp}}(\langle A_J \rangle),
\end{equation}
where $N_{\mathrm{exp}}(A_J)$ is computed by assuming a uniform extinction across the cell, and the average is taken over the distribution of $A_J$ values from the random catalogue.

This comparison was performed across a large number of cells spanning different regions of the sky, including those close to the Galactic plane where both the extinction and its spatial variation are largest. 
In all cases, the two quantities agree to better than $\sim 0.1\%$, while the spread in $N_{\mathrm{exp}}(A_J)$ across the extinction distribution remains at only the few percent level. 
This is much smaller than the real cell-to-cell luminosity-density variations that are the primary focus of this work. 
Figure~\ref{fig:dust_linearity} shows an example from one of the most affected cells, located close to the Zone of Avoidance.

This behaviour indicates that the dependence of the expected counts on extinction is close to linear over the range present in each cell. 
As a result, replacing the full distribution of $A_J$ with its mean provides an accurate approximation. 
This remains true even in regions of high extinction.

\section{Finger-of-God Effects}
\label{appendix:fog}

\begin{table*}
\centering
\caption{Comparison of integrated luminosities with and without approximate Finger-of-God corrections for several nearby galaxy clusters. The luminosities are computed using the summed galaxy luminosities within the selected cluster regions.}
\label{tab:fog_clusters}
\begin{tabular}{lccccc}
\hline
Cluster & Number of Galaxies & $\log_{10}(L_{\rm no\,FoG})$ & $\log_{10}(L_{\rm FoG})$ & $\Delta \log_{10}(L)$ & $L_{\rm FoG}/L_{\rm no\,FoG}$ \\
 &  & [$L_{\odot}$] & [$L_{\odot}$] & [dex] &  \\
\hline
Coma Cluster      & 180 & 13.049 & 13.047 & -0.003 & 0.994 \\
Hydra Cluster     & 283 & 12.874 & 12.884 &  0.009 & 1.021 \\
Centaurus Cluster & 285 & 12.872 & 12.856 & -0.017 & 0.962 \\
Virgo Cluster     &  45 & 11.808 & 11.454 & -0.354 & 0.443 \\
\hline
\end{tabular}
\end{table*}

\begin{figure*}
    \centering
    \includegraphics[width=\textwidth]{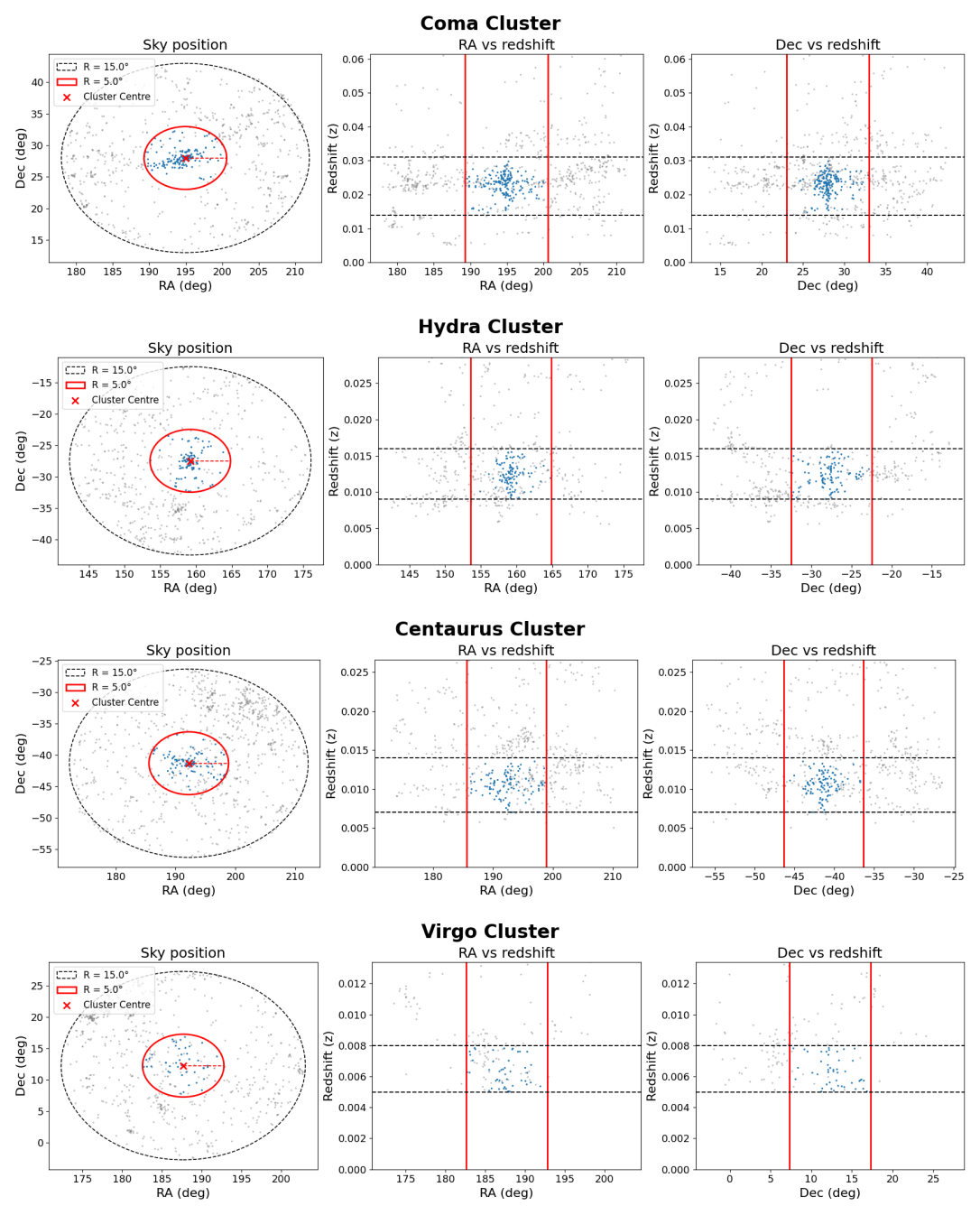}
    \caption{Selected nearby galaxy clusters used for the Finger-of-God analysis. 
    The left panels show the sky selection regions, while the middle and right panels show the corresponding redshift distributions as a function of right ascension and declination respectively. 
    The red boundaries indicate the adopted cluster selection regions used in the FoG correction test.}
    \label{fig:fog_clusters}
\end{figure*}

One potential source of systematic error in low-redshift luminosity function studies is the presence of Finger-of-God (FoG) distortions within nearby galaxy clusters. 
Galaxies inside virialised clusters possess significant peculiar velocities relative to the cluster centre, causing their observed redshifts to differ from the true cosmological redshift associated with their physical distance. 
If interpreted directly as distance indicators, these peculiar velocities can artificially broaden the inferred absolute magnitude distribution and potentially bias the recovered luminosity function and luminosity density measurements.

To estimate the magnitude of this effect within the present analysis, we perform a simple FoG correction test on several nearby and prominent structures within the 2MRS sample. 
The Coma, Hydra, Centaurus and Virgo clusters were selected as representative systems, as these nearby overdensities are expected to produce the strongest local FoG signatures in the survey volume. 
For each cluster, galaxies were selected within an angular aperture and redshift range centred on the cluster. 
We then compare two cases:

\begin{enumerate}
    \item using the observed galaxy redshifts directly when computing luminosity distances, and
    \item assigning all galaxies within the selected region the mean cluster redshift when computing luminosity distances.
\end{enumerate}

The second case approximately removes the line-of-sight velocity dispersion contribution and therefore provides a simple estimate of the maximum FoG-induced luminosity shift within the cluster core.

Figure~\ref{fig:fog_clusters} shows the selected cluster regions and corresponding redshift distributions used in this analysis. 
For each system, we compute the total integrated luminosity by summing all the magnitudes,

\begin{equation}
    \log_{10}\left(\sum_i L_i\right),
\end{equation}

where the luminosity of each galaxy is calculated using

\begin{equation}
    L_i = 10^{-0.4(M_i - M_{\odot,J})},
\end{equation}

with $M_{\odot,J}=3.65$ in the Vega system.

The resulting luminosity differences are summarised in Table~\ref{tab:fog_clusters}. 
For the Coma, Hydra and Centaurus clusters, the inferred changes in total luminosity are very small, with differences below the percent level in all cases. 
These systems are relatively compact and dynamically relaxed compared to Virgo, and the FoG correct ion therefore produces only a negligible change in the integrated luminosity budget.

Virgo, however, exhibits a substantially larger shift. 
Unlike the other systems considered here, Virgo is a significantly more complex and extended structure and is not virialised in the same manner as Coma or Hydra.
Due to its proximity, Virgo spans a large angular region on the sky, and defining cluster membership using only simple positional and redshift cuts becomes substantially more difficult. 
As a result, the present correction likely includes galaxies that are not physically associated with the Virgo core, artificially amplifying the inferred FoG correction. 
This illustrates an important practical point: an imperfect group or cluster correction can itself introduce luminosity-density errors that are comparable to, or larger than, the FoG effect it is intended to remove. 
For this reason, applying group-based corrections without a uniform and reliable group catalogue may not improve the measurement, and can potentially make the inferred luminosity density less robust.

The primary conclusion from this test is therefore that outside of Virgo's center, which is both extremely nearby and dynamically complex, FoG effects are comparitively small for the broader nearby cluster populations.
While a more complete treatment of peculiar velocity corrections will be incorporated in future work, the present analysis suggests that FoG distortions are unlikely to introduce a significant systematic bias into the global luminosity density measurements presented in this work.


\bsp	
\label{lastpage}
\end{document}